\documentclass[12pt]{iopart}

\usepackage{iopams}
\usepackage{graphicx}
\usepackage{pseudocode}
 \expandafter\let\csname equation*\endcsname\relax
  \expandafter\let\csname endequation*\endcsname\relax
\usepackage{amsmath}
\usepackage{amssymb}
\usepackage{amsthm}  
\usepackage{color}

\newcommand{\vm}[1]{\boldsymbol{#1}}
\renewcommand{\thepseudocode}{1}

\begin{document}

\title{Statistical mechanics approach to 1-bit compressed sensing}
\author{Yingying Xu and Yoshiyuki Kabashima }

\address{Department of Computational Intelligence and Systems Science, 
\\Tokyo Institute of Technology, Yokohama 226-8502, Japan}
\ead{yingxu@sp.dis.titech.ac.jp, kaba@dis.titech.ac.jp}

\begin{abstract}
Compressed sensing is a framework that makes it possible to recover an $N$-dimensional sparse vector $\textrm{\boldmath $x$}\in \mathbb{R}^N$ from its linear transformation $\textrm{\boldmath $y$}\in \mathbb{R}^M$ of lower dimensionality $M < N$. A scheme further reducing the data size of the compressed expression by using only the sign of each entry of $\textrm{\boldmath $y$}$ to recover $\textrm{\boldmath $x$}$ was recently proposed. This is often termed the {\em 1-bit compressed sensing}. Here we analyze the typical performance of an $l_1$-norm based signal recovery scheme for the 1-bit compressed sensing using statistical mechanics methods. We show that the signal recovery performance predicted by the replica method under the replica symmetric ansatz, which turns out to be locally unstable for modes breaking the replica symmetry, is in a good consistency with experimental results of an approximate recovery algorithm developed earlier. This suggests that the $l_1$-based recovery problem typically has many local optima of a similar recovery accuracy, which can be achieved by the approximate algorithm. We also develop another approximate recovery algorithm inspired by the cavity method. Numerical experiments show that when the density of nonzero entries in the original signal is relatively large the new algorithm offers better performance than the abovementioned scheme and does so with a lower computational cost. 
\end{abstract}

\maketitle

\section{Introduction}
\textit{Compressed (or compressive) sensing} (CS) is a technique for recovering a high-dimensional signal from lower-dimensional data, whose components represent partial information about the signal, by utilizing prior knowledge on the sparsity of the signal \cite{CandesWakin2008}. The research field of CS is one of the main topics in information science nowadays and has been intensively investigated from the theoretical point of view \cite{Donoho2006, CandesRombergTao2006,KWT2009,Sompolinsky2010,Krzakala2012}. This technique has also been used in various engineering fields \cite{CShardware}.

Let us suppose a situation that an $N$-dimensional vector $\textrm{\boldmath $x^{0}$}$ is linearly transformed into an $M$-dimensional vector $\textrm{\boldmath $y$}$ by an $M \times N$ measurement matrix $\boldmath \Phi$, where $\textrm{\boldmath $y$}=\textrm{\boldmath $\Phi x^{0}$}$. The signal is recovered from $\textrm{\boldmath $y$}$ by determining the sparsest signal that is consistent with the measurements. When $M < N$, the measurement usually loses some information and the inverse problem has an infinite number of solutions. However, when the $N$-dimensional signal is guaranteed to have only $K<M$ nonzero entries in some convenient basis and the measurement matrix is {incoherent} with that basis, there is a high probability that the inverse problem has a unique and exact solution. For example, smooth signals and piecewise-smooth signals, like natural images or communications signals, typically have a representation in a sparsity-inducing basis such as a Fourier or wavelet basis \cite{Elad2010,StarckMurtaghFadili2010}. 

Although most of the compressed sensing literature has not explicitly handled quantization of the measured data until recently, quantizing continuous data is unavoidable in most real-world applications, particularly those in which the measurement is accompanied by digital information transmission \cite{Lee2012}. Addressing the practical relevance of CS in such operation, Boufounos and Baraniuk recently proposed and examined a CS scheme, often called {\em 1-bit compressed sensing (1-bit CS)}, in which the signal is recovered from only the sign data of the linear measurements $\textrm{\boldmath $y$}=\mathrm{sign}\left(\textrm{\boldmath $\Phi x^{0}$}\right)$, where $\mathrm{sign}(x)=x/|x|$ for $x \ne 0$ operates for vectors in the component-wise manner \cite{1bitCS}. Although in 1-bit CS the amplitude information is lost during the measurement stage, making perfect recovery of the original signal impossible, discarding the amplitude information can significantly reduce the amount of data that needs to be stored and/or transmitted. This is highly advantageous when perfect recovery is not required. In addition, quantization to the 1-bit (sign) information is appealing in hardware implementations because 1-bit quantizer takes the form of a comparator to zero and does not suffer from dynamic range issues. The scheme is considered practical relevant in situations where measurements are inexpensive and precise quantization is expensive, in which the cost of measurements should be quantified by the number of total bits needed to store the data instead of by the number of measurements. 

The purpose of this paper is to explore the abilities and limitations of a 1-bit CS scheme utilizing statistical mechanics methods. In \cite{1bitCS} an approximate signal recovery algorithm based on minimization of the $l_1$-norm $||\textrm{\boldmath $x$}||_1=\sum_{i=1}^N |x_i|$ under the constraint of $\mathrm{sign}\left(\textrm{\boldmath $\Phi x$}\right)=\textrm{\boldmath $y$}$ was proposed and its utility was shown by numerical experiments. Quantization to the sign information, however, leads to the loss of the convexity of the resulting optimization problem, which makes it difficult to mathematically examine how well the obtained solution approximates the correct solution. Comparing (in terms of the mean square error) the results of numerical experiments with the theoretical prediction evaluated by the replica method \cite{replica}, we will show that the performance of the approximate algorithm is nearly as good as %[?approaches?] 
that potentially achievable by the $l_1$-based scheme. We will also develop another approximate algorithm inspired by the cavity method \cite{cavity,MezardMontanari2009} and will show that when the density of nonzero entries of the original signal is relatively high the new algorithm offers better recovery performance with much lower computational cost. 

This paper is organized as follows. The next section sets up the problem that we will focus on when explaining the 1-bit CS scheme. Section 3 uses the replica method to examine the signal recovery performance achieved by the scheme. In section 4 an approximate signal recovery algorithm based on the cavity method is developed and evaluated, and the final section is devoted to a summary. 

\section{Problem setup}
Let us suppose that entry $x_i^0$ $(i=1,2,\ldots,N)$ of $N$-dimensional signal (vector) $\vm{x}^0 \in \mathbb{R}^N$ is independently generated from an identical sparse distribution:
\begin{equation}
P\left(x\right)=\left(1-\rho\right) \delta \left( x \right)
+ \rho \tilde{P} \left( x \right), 
\label{sparse}
\end{equation}
where $\rho \in [0,1]$ represents the density of nonzero entries in the signal and $\tilde{P} (x)$ is a distribution function of $x \in \mathbb{R}$ that does not have finite mass at $x=0$. In 1-bit CS the measurement is performed as 
\begin{equation}
\textrm{\boldmath $y$}=\mathrm{sign} \left( \textrm{\boldmath $\Phi x^{0}$} \right), 
\label{measurement}
\end{equation}
where for simplicity we assume that each entry of $M \times N$ measurement matrix $\vm{\Phi}$ is provided as an independent sample from an identical Gaussian distribution of zero mean and variance $N^{-1}$. 

Given $\vm{y}$ and $\vm{\Phi}$, the signal reconstruction is carried out by searching for a sparse vector $\vm{x} =(x_i) \in \mathbb{R}^N$ under the constraint of ${\rm sign}\left (\vm{\Phi} \vm{x} \right )=\vm{y}$. For this task the authors of \cite{1bitCS} proposed a scheme of 
\begin{eqnarray}
\mathop{\rm min}_{\vm{x} } \left \{||\vm{x}||_1 \right \} \ {\rm subj. \ to} \ 
{\rm sign}\left (\vm{\Phi} \vm{x} \right )=\vm{y} \ {\rm and} \
||\vm{x}||_2=\sqrt{N}, 
\label{l1recovery}
\end{eqnarray}
based on the $l_1$-recovery method widely used and studied for standard CS problems \cite{CandesWakin2008}. Here $|| \vm{x}||_1 =\sum_{i=1}^N |x_i|$ and $||\vm{x}||_2=|\vm{x}|=\sqrt{\sum_{i=1}^N x_i^2}$ denote the $l_1$- and $l_2$-norms of $\vm{x}$, respectively. The measurement process of (\ref{measurement}) completely erases the information of length $|\vm{x}^0|$, which makes it impossible to recover the signal uniquely. We therefore introduce an extra normalization constraint $|\hat{\vm{x}}|=\sqrt{N}$ for the recovered signal $\hat{\vm{x}}$, and we consider the recovery successful when the direction cosine ${\vm{x}}^0 \cdot \hat{\vm{x}}/(|\vm{x}^0| |\hat{\vm{x}}|)$ is sufficiently large. 

\begin{figure}
  \begin{center}
   \includegraphics[width=14cm]{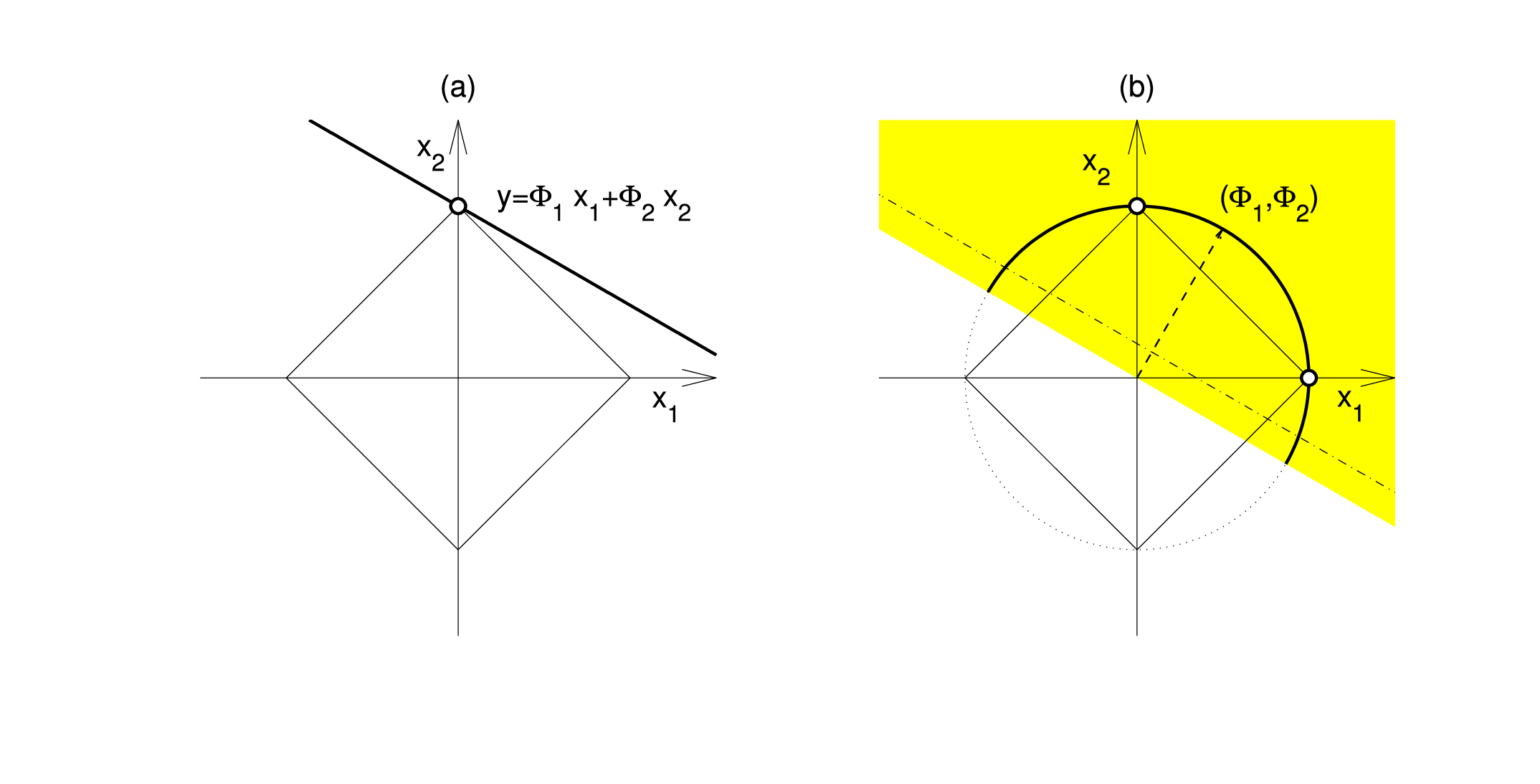}
  \end{center}
  \vspace*{-1.5cm}
\caption{\protect\label{Normal_vs_1bit}Graphical representations of (a) standard and (b) 1-bit CS problems in the case of $N=2$, $M=1$, and $K=\rho N=1$. (a): A thick line and a square of thin lines represent a measurement result $y=\Phi_1 x_1+\Phi_2 x_2$ and a contour of $l_1$-norm $|x_1|+|x_2|$, respectively. The optimal solution denoted by a circle is uniquely determined since both the set of feasible solutions $y=\Phi_1 x_1+\Phi_2 x_2$ and the cost function $|x_1|+|x_2|$ are convex. (b): The shaded area $y\times (\Phi_1 x_1+\Phi_2 x_2)>0$ represents the region that is compatible with the sign information of the linear measurement $y=\Phi_1 x_1+\Phi_2 x_2$ (dotted broken line). This and the $l_2$-norm constraint $x_1^2+x_2^2=2$ yield the set of feasible solutions as a semicircle (thick curve), which is not a convex set. As a consequence, the constraint optimization problem of (\ref{l1recovery}) generally has multiple solutions (two circles).}
\end{figure}

Unlike the standard CS problem, finding a solution of (\ref{l1recovery}) is non-trivial because the norm constraint $|{\vm{x}}|=\sqrt{N}$ keeps it from being a convex optimization problem (Figures \ref{Normal_vs_1bit} (a) and (b)). The authors of \cite{1bitCS} also developed, as a practically feasible solution, a double-loop algorithm called Renormalized Fixed Point Iteration (RFPI) that combines a gradient descent method and enforcement to a sphere of a fixed radius. It is summarized in Figure \ref{RFPI}. 

The practical utility of RFPI was shown by numerical experiments, but 
how good solutions are actually obtained is unclear 
because in general 
the algorithm can be trapped at various local optima. 
One of our main concerns is therefore to theoretically evaluate the typical performance of the global minimum solution of (\ref{l1recovery})
for examining the possibility of performance improvement. 

\begin{figure}
\begin{pseudocode}[ruled]{Renormalized Fixed Point Iteration}{\delta,\lambda}
1)\ \mbox{\bf Initialization}:\\
 \hspace{15pt}\text{Seed}: \hspace{80pt}\hat{\mathbf{x}}_{0}\ \text{s.t}\ ||\hat{\mathbf{x}}_{0}||_{2}=\sqrt{N},\\
 \hspace{15pt}\text{Descent step size}:\hspace{13pt}\delta\\
\hspace{15pt}\text{Counter}: \hspace{62pt} k\GETS 0\\

2)\ \mbox{\bf Counter\ Increase}:\\
 \hspace{30pt}k \GETS k+1\\

3)\ \mbox{\bf One-sided quadratic gradient}:\\
\hspace{30pt}\mathbf{\overline{f}}_{k}\GETS  (\textsc{Y}\Phi)^{\rm T} f^\prime (\textsc{Y}\Phi \hat{\mathbf{x}}_{k-1})\\

4)\ \mbox{\bf Gradient projection on sphere surface}:\\
 \hspace{30pt}\mathbf{\tilde{f}}_{k}\GETS \mathbf{\overline{f}}_{k}-\langle \mathbf{\overline{f}}_{k},
 \hat{\mathbf{x}}_{k-1}\rangle\hat{\mathbf{x}}_{k-1}/N\\

5)\ \mbox{\bf One-sided quadratic gradient descent}:\\
 \hspace{30pt}\mathbf{h}\GETS \hat{\mathbf{x}}_{k-1}-\delta\mathbf{\tilde{f}}_{k}\\

6)\ \mbox{\bf Shrinkage ($l_1$-gradient descent)}:\\ 
 \hspace{30pt}(\mathbf{u})_i\GETS\text{sign}((\mathbf{h})_{i})\text{max}\{|(\mathbf{h})_i|-\frac{\delta}{\lambda},0\} \ \text{for all}\ i \\

7)\ \mbox{\bf Normalization}:\\
\hspace{30pt}\hat{\mathbf{x}}_{k}\GETS \sqrt{N} \frac{\mathbf{u}}{||\mathbf{u}||_2}\\

8)\ \bold{Iteration}: \mbox{Repeat from 2) until convergence.}
\end{pseudocode}
\protect
\caption{\protect\label{RFPI}Pseudocode for the inner loop of the Renormalized Fixed Point Iteration (RFPI) proposed in \cite{1bitCS}. The function $f^\prime (x)$ in step 3 is defined as $f^\prime (x)=x$ for $x \le 0$ and $0$, otherwise, and it operates on a vector in a component-wise manner. In the original expression in \cite{1bitCS} the normalization constraint is introduced as $||\hat{\mathbf{x}}_{k}||_2=1$, but we here use 
$||\hat{\mathbf{x}}_{k}||_2=\sqrt{N}$ for convenience in considering the large system limit of $N \to \infty$. RFPI is a double-loop algorithm. In the outer loop the parameter $\lambda$ is increased as $\lambda_n=c \lambda_{n-1}$, where $c>1$ and $n$ are a certain constant and the counter of the outer loop, respectively. The convergent solution of $i-1$th outer loop is used for the initial state of the inner loop of the $i$th outer loop. The algorithm terminates when difference between the convergent solutions of two successive outer loops become sufficiently small.}
\end{figure}

\section{Performance assessment by the replica method}
The partition function 
\begin{equation}
Z\left(\beta; \Phi, \textrm{\boldmath $x$}^{0} \right)=\int d \textrm{\boldmath $x$}
\delta\left(\vert \textrm{\boldmath $x$}\vert^{2}-N\right)e^{-\beta\vert\vert\textrm{\boldmath $x$}\vert\vert _{1}}
\prod_{\mu=1}^M \Theta\left ( (\vm{\Phi} \vm{x}^0 )_\mu  (\vm{\Phi} \vm{x})_\mu \right )
\label{eq:partition_func},
\end{equation}
where $\Theta\left(x \right)=1$ 
%% kaba 12/12/12
and $0$ for $x>0$ and $x<0$, respectively, offers the basis for our analysis. As $\beta$ tends to infinity, the integral of (\ref{eq:partition_func}) is dominated by the correct solution of (\ref{l1recovery}). One therefore can evaluate the performance of the solution by examining the macroscopic behavior of equation (\ref{eq:partition_func}) in the limit of $\beta\rightarrow \infty$.

A characteristic feature of the current problem is that (\ref{eq:partition_func}) depends on the predetermined random variables $\textrm{\boldmath $\Phi$}$ and $\textrm{\boldmath $x^{0}$} $, which requires us to assess the average of free energy density $f \equiv -(\beta N)^{-1} \left [\ln Z(\beta; \vm{\Phi},\vm{x}^0) \right]_{\vm{\Phi},\vm{x}^0}$ when evaluating the performance for typical samples of $\vm{\Phi}$ and $\vm{x}^0$. Here, $\left [\cdots \right]_{\vm{\Phi},\vm{x}^0}$ denotes the configurational average concerning $\textrm{\boldmath $\Phi$}$ and $\textrm{\boldmath $x^{0}$}$. Because directly averaging the logarithm of the partition function is technically difficult, we here resort to the replica method \cite{replica}. 

For this we first evaluate $n$-th moment of the partition function $\left [Z^n \left(\beta; \vm{\Phi}, \textrm{\boldmath $x$}^{0} \right) \right]_{\vm{\Phi},\vm{x}^0}$ for $n=1,2,\ldots \in \mathbb{N}$, using the formula 
\begin{eqnarray}
&&Z^n \left(\beta; \vm{\Phi}, \textrm{\boldmath $x$}^{0} \right)
=
\int \prod_{a=1}^n \left (d \textrm{\boldmath $x$}^a
\delta\left(\vert \textrm{\boldmath $x$}^a\vert^{2}-N\right) \times
e^{-\beta\vert\vert\textrm{\boldmath $x^a$}\vert\vert _{1}}  \right )\cr
&& \hspace*{4cm} \times 
\prod_{a=1}^n \prod_{\mu=1}^M \Theta\left ( (\vm{\Phi} \vm{x}^0 )_\mu  (\vm{\Phi} \vm{x}^a)_\mu \right ), 
\label{eq:expansion}
\end{eqnarray}
which holds only for $n=1,2,\ldots \in \mathbb{N}$. 
%% kaba 12/12/12
Here, $\vm{x}^a$ ($a=1,2,\ldots,n$) denotes  $a$-th replicated signal. 
Averaging (\ref{eq:expansion}) with respect to $\vm{\Phi}$ and $\vm{x}^0$ results in the saddle point evaluation concerning macroscopic variables $q_{0a} =q_{a0}\equiv N^{-1} \vm{x}^0 \cdot \vm{x}^a$ and $q_{ab}=q_{ba} \equiv N^{-1} \vm{x}^a \cdot \vm{x}^b$ ($a,b =0,1,2,\ldots,n$). % concerning $\vm{x}^0$ and $n$ replicated signals $\vm{x}^1,\vm{x}^2,\ldots,\vm{x}^n$. 
Although (\ref{eq:expansion}) holds only for $n \in \mathbb{N}$, the expression of $N^{-1} \ln \left [Z^n \left(\beta; \vm{\Phi}, \textrm{\boldmath $x$}^{0} \right) \right]_{\vm{\Phi},\vm{x}^0}$ obtained by the saddle point evaluation under a certain assumption concerning the permutation symmetry with respect to the replica indices $a,b=1,2,\ldots n$ is obtained as an analytic function of $n$, which is likely to also hold for $n \in \mathbb{R}$. Therefore, we next utilize the analytic function for evaluating the average of the logarithm of the partition function as $N^{-1} \ln \left [\ln Z(\beta; \vm{\Phi},\vm{x}^0) \right]_{\vm{\Phi},\vm{x}^0} \lim_{n \to 0} N^{-1} \ln \left [Z^n \left(\beta; \vm{\Phi}, \textrm{\boldmath $x$}^{0} \right) \right]_{\vm{\Phi},\vm{x}^0}$. 

In particular, under the replica symmetric (RS) ansatz where the dominant saddle point is assumed to be of the form of 
\begin{eqnarray}
q_{ab}=q_{ba}=\left \{
\begin{array}{ll}
\rho & (a=b=0) \cr
m & (a=1,2,\ldots,n; \ b=0) \cr
1 & (a=b=1,2,\ldots,n) \cr
q & (a\ne b =1,2,\ldots,n) 
\end{array}
\right . ,  
\label{RSanzats}
\end{eqnarray}
when the distribution of nonzero entries in (\ref{sparse}) is given as the standard Gaussian $\tilde{P}(x)=\exp(-x^2/2)/\sqrt{2 \pi}$, the above procedure offers an expression of the average free energy density as 
\begin{eqnarray}
\bar{f} 
  &=& \mathop{\rm extr}_{\omega} \Biggr\{ \left[\phi \left(\sqrt{\hat{q}}z+\hat{m}\textrm{\boldmath $x$}^{0};\hat{Q}\right)\right]_{\textrm{\boldmath $x$}^{0},z}
-\frac{1}{2}\hat{Q}+\frac{1}{2}\hat{q}\chi+\hat{m}m \nonumber\\
 &&+\frac{\alpha}{2\pi\chi}\left(\arctan \left (\frac{\sqrt{\rho-m^2}}{m} \right )-\frac{m}{\rho}\sqrt{\rho-m^2}\right) \Biggr\} 
\label{eq:free energy}
\end{eqnarray}
in the limit of $\beta \to \infty$. Here $\alpha=M/N$, $\textrm{extr}_{X}\{g(X)\}$ denotes extremization of a function $g(X)$ with respect to $X$, $\omega = \{\chi, m, \hat{Q}, \hat{q}, \hat{m}\}$, $\textrm{D}z=\textrm{d}z \textrm{exp}(-z^2/2)/\sqrt{2\pi}$ is a Gaussian measure, and
\begin{eqnarray}
\phi \left(\sqrt{\hat{q}}z+\hat{m}\textrm{\boldmath $x$}^{0};\hat{Q}\right)
=\mathop{\rm min}_{x}\left\{ \frac{\hat{Q}}{2}x^2-\left(\sqrt{\hat{q}}z+\hat{m}x^0\right)x+\vert x \vert\right\}\nonumber\\
= -\frac{1}{2\hat{Q}}\left( \left \vert\sqrt{\hat{q}}z
+\hat{m}\textrm{\boldmath $x$}^{0} \right \vert-1\right)^2
\Theta \left (\left \vert\sqrt{\hat{q}}z+\hat{m}\textrm{\boldmath $x$}^{0} \right \vert-1 \right ).
\end{eqnarray}
The derivation of $(\ref{eq:free energy})$ is provided in \ref{replicaderivation}.

The extremization problem of (\ref{eq:free energy}) yields the following saddle point equations: 
\begin{eqnarray}
\hat{q}&=&\frac{\alpha}{\pi \chi^2} \left(\arctan \left(\frac{\sqrt{\rho-m^2}}{m}\right) -\frac{m}{\rho}\sqrt{\rho-m^2}\right), \\
\hat{m}&=&\frac{\alpha}{\pi\chi\rho}\sqrt{\rho-m^2}, \\
\hat{Q}^{2}&=&2\left\{ \left(1-\rho\right)\left[\left(\hat{q}+1\right)H\left(\frac{1}{\sqrt{\hat{q}}}\right)
          -\sqrt{\frac{\hat{q}}{2\pi}}e^{-\frac{1}{2\hat{q}}} \right] \right . \nonumber\\
          &&\left . +\rho\left[\left(\hat{q}+\hat{m}^2+1\right)H\left(\frac{1}{\sqrt{\hat{q}+\hat{m}^2}}\right)
          -\sqrt{\frac{\hat{q}+\hat{m}^2}{2\pi}}e^{-\frac{1}{2\left(\hat{q}+\hat{m}^2\right)}}\right]\right\}, \\
\chi&=&\frac{2}{\hat{Q}}\left[\left(1-\rho\right)
H\left(\frac{1}{\sqrt{\hat{q}}}\right)+\rho H\left(\frac{1}{\sqrt{\hat{q}+\hat{m}^2}}\right)\right], \\
m&=&\frac{2\rho\hat{m}}{\hat{Q}}H\left(\frac{1}{\sqrt{\hat{q}+\hat{m}^2}}\right), 
\label{results}
\end{eqnarray}
where $H(x)=\int_x^{+\infty} {\rm D}z$. The value of $m$ determined by these equations physically means the typical overlap $N^{-1} \left [\vm{x}^0 \cdot \hat{\vm{x}} \right]_{\vm{\Phi}, \vm{x}^0} $ between the original signal $\vm{x}^0$ and the solution $\hat{\vm{x}}$ of (\ref{l1recovery}). Therefore the typical value of the direction cosine between $\vm{x}^0$ and $\hat{\vm{x}}$, which serves as a performance measure of the current recovery problem, is evaluated as $\left [(\vm{x}^0 \cdot \hat{\vm{x}})/|\vm{x}^0| |\hat{\vm{x}}| \right]_{\vm{\Phi,\vm{x}^0}} =N m /(\sqrt{N \rho} \times \sqrt{N})=m/\sqrt{\rho}$. Alternatively, we may also use as a performance measure the mean square error (MSE) between the normalized vectors:
\begin{eqnarray}
{\rm MSE}=\left [ \left |\frac{\hat{\vm{x}}}{|\hat{\vm{x}}|}-\frac{\vm{x}^0}{|\vm{x}^0|} \right |^2 \right ]_{\vm{\Phi,\vm{x}^0}}
=2\left (1-\frac{m}{\sqrt{\rho}} \right ).
\label{MSE}
\end{eqnarray}

\begin{figure}[t]
  \begin{center}
    \begin{tabular}{c}
    
      \begin{minipage}{0.5\hsize}
        \begin{center}
          \includegraphics[clip,angle=270,width=8.5cm]{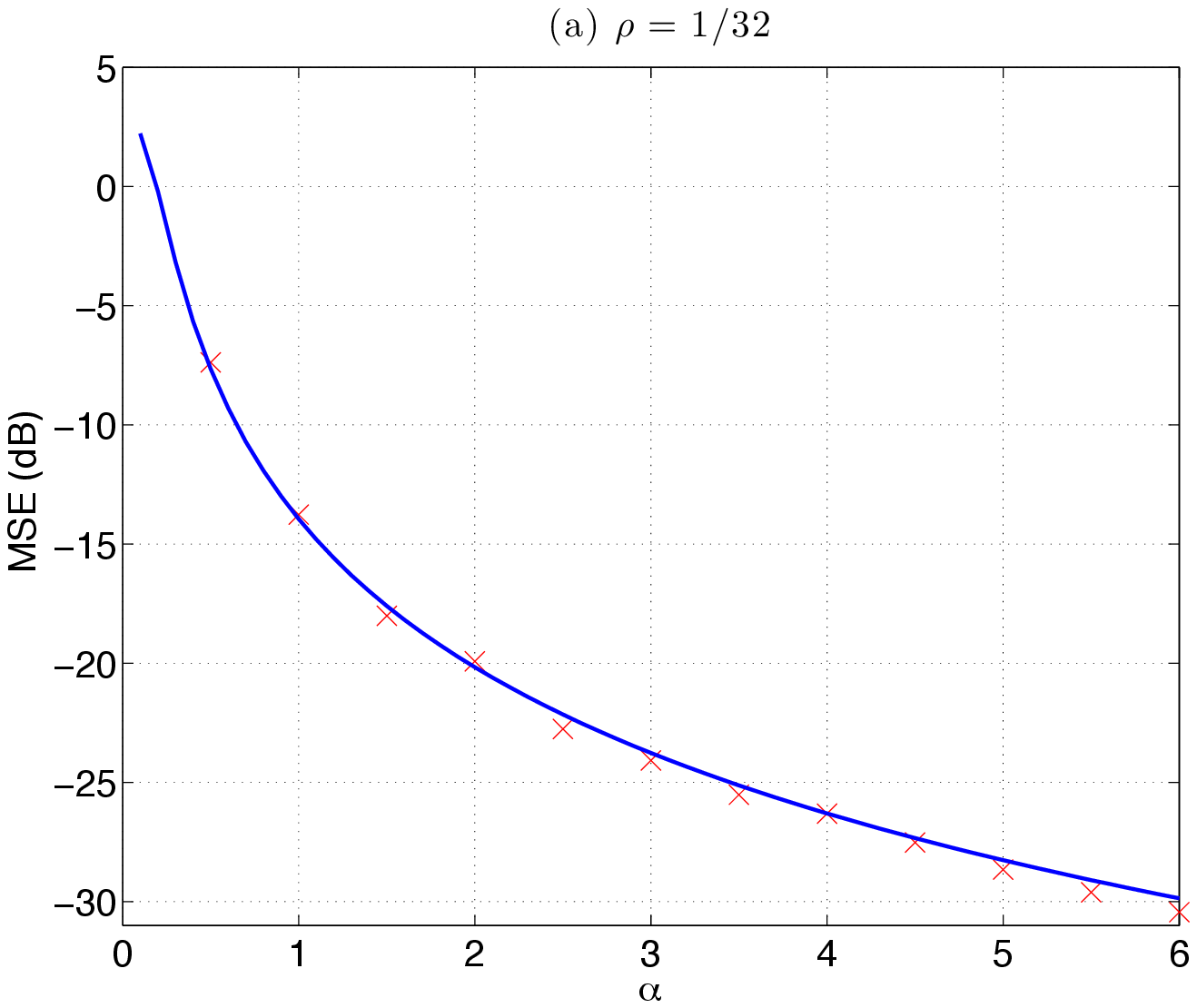}
          %\hspace{1.6cm} 
        \end{center}
      \end{minipage}
      
      \begin{minipage}{0.5\hsize}
        \begin{center}
          \includegraphics[clip,angle=270,width=8.5cm]{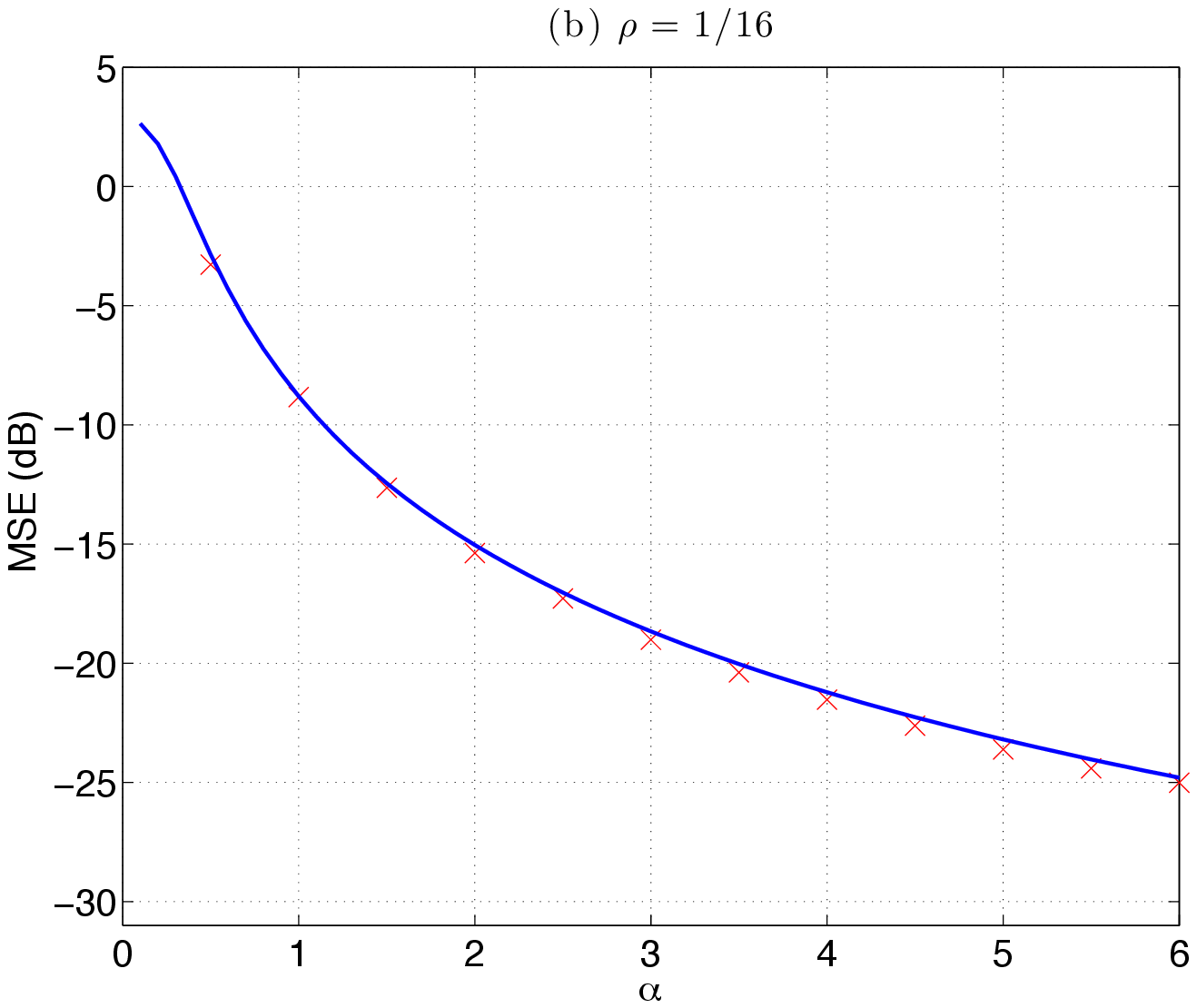}
          %\hspace{1.6cm}
        \end{center}
      \end{minipage}
\\
      \begin{minipage}{0.5\hsize}
        \begin{center}
          \includegraphics[clip, angle=270,width=8.5cm]{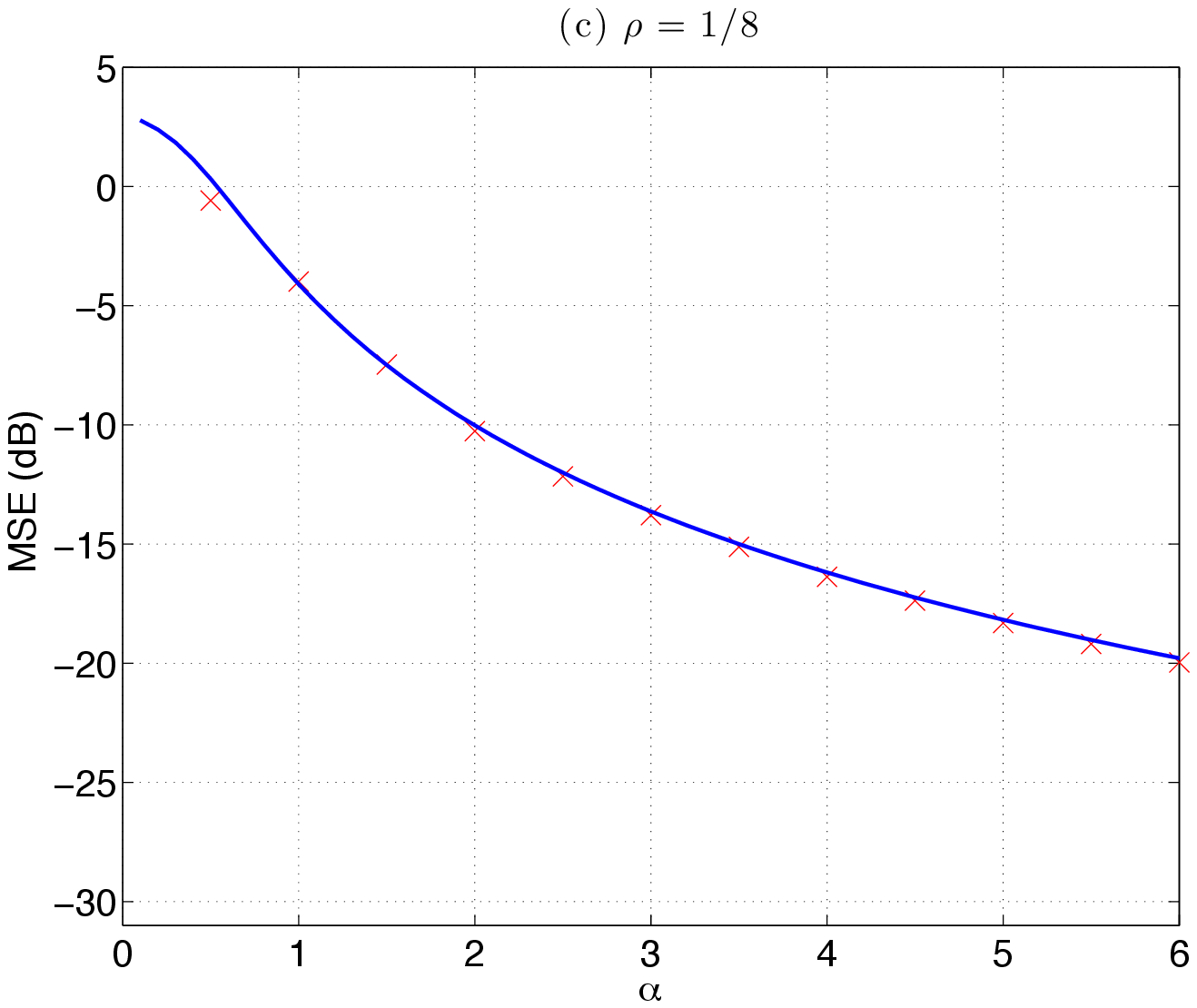}
          %\hspace{1.6cm} 
        \end{center}
      \end{minipage}
      
      \begin{minipage}{0.5\hsize}
        \begin{center}
          \includegraphics[clip, angle=270,width=8.5cm]{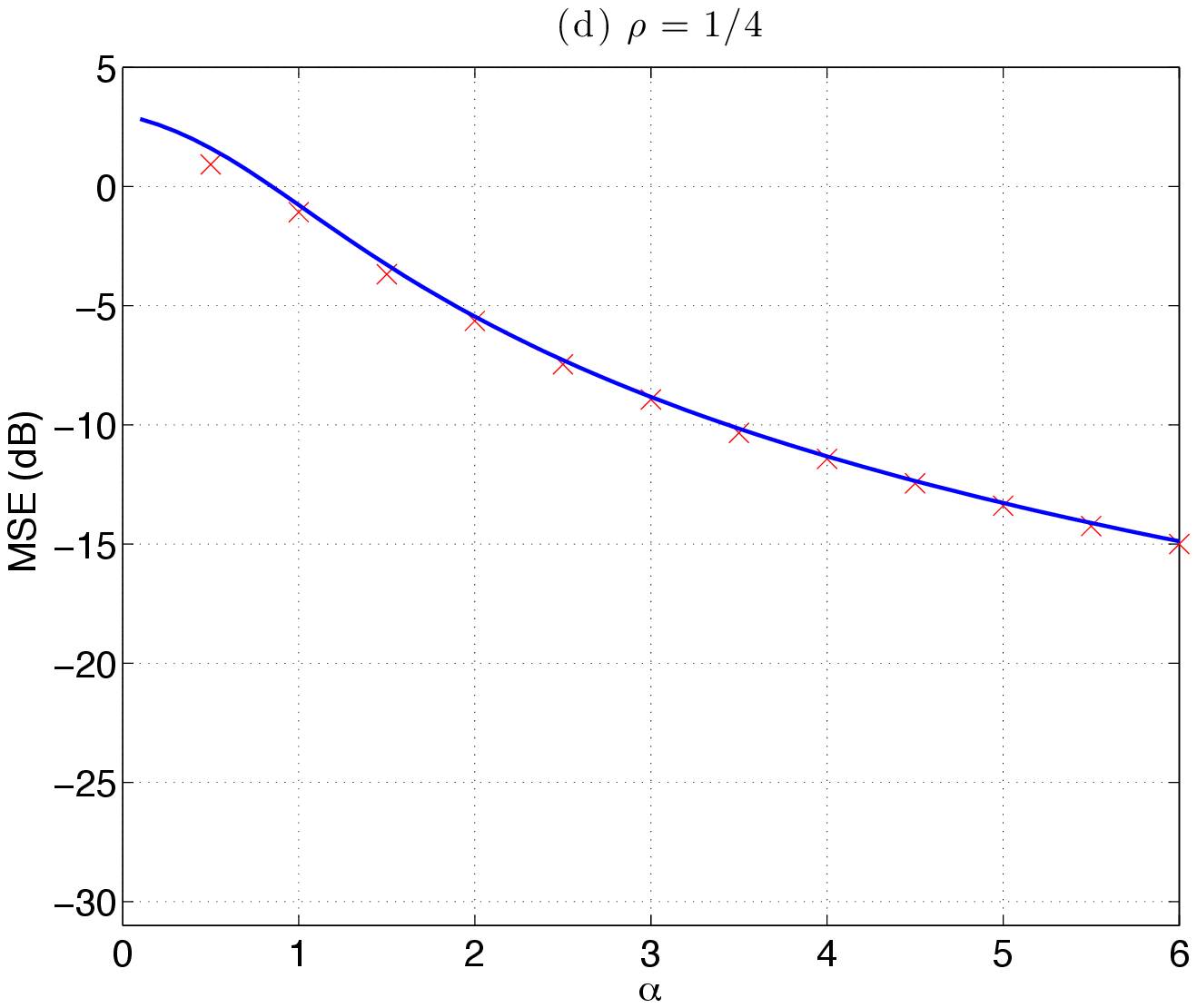}
          %\hspace{1.6cm} 
        \end{center}
      \end{minipage}
    \end{tabular}
\caption{\label{fig2}MSE versus the measurement bit ratio $\alpha$ for the signal recovery scheme using (\ref{l1recovery}). (a), (b), (c), and (d) correspond to the $\rho=1/32, 1/16, 1/8$, and $1/4$ cases, respectively. Curves represent the theoretical prediction evaluated by the RS solution, which is locally unstable for disturbances that break the replica symmetry for all regions of (a)--(d). Each symbol ($\times$) stands for the experimental estimate obtained for RFPI in \cite{1bitCS} from $1000$ experiments with $N=128$ systems.}
%[In the axis labels here,  MSE(dB)  -->  MSE (dB)]

\end{center}
\end{figure}

We solved the saddle point equations for various sets of $\alpha$ and $\rho$. The curves in figures \ref{fig2} (a)--(d) show the theoretical prediction of MSE evaluated by (\ref{MSE}) plotted against the measurement bit ratio $\alpha=M/N$ for $\rho=1/32, 1/16, 1/8$, and $1/4$. To examine the validity of the RS ansatz, we also evaluated the local stability of the RS solutions against the disturbances that break the replica symmetry \cite{AT}, which offers 
\begin{eqnarray}
&&\frac{\alpha}{\pi (\hat{Q}\chi)^2} \arctan \left (\frac{\sqrt{\rho-m^2}}{m} \right ) \cr
&& \times 2\left ((1-\rho)H\left (\frac{1}{\sqrt{\hat{q}}} \right )
+\rho H\left (\frac{1}{\sqrt{\hat{q}+\hat{m}^2}} \right ) \right ) -1<0, 
\label{AT}
\end{eqnarray}
as the stability condition. A brief sketch of the derivation of this condition is shown in \ref{RSstability}. Unfortunately, (\ref{AT}) is not satisfied for any regions in figures \ref{fig2} (a)--(d). This is presumably because the optimization problem for (\ref{l1recovery}) has many local optima reflecting the fact that the constraint of $||\vm{x}||_2=\sqrt{N}$ loses the convexity. This indicates that taking the replica symmetry breaking (RSB) into account is necessary for evaluating the exact performance of the signal recovery scheme defined by (\ref{l1recovery}). 

We nonetheless think that the RS analysis offers considerably accurate approximates of the exact performance in terms of MSE. The ($\times$) symbols in figures \ref{fig2} (a)--(d) stand for MSE experimentally achieved by RFPI, which were assessed as the arithmetic averages over $1000$ samples for each condition of $N=128$ systems. Excellent consistency between the curves and symbols suggests that even if (\ref{l1recovery}) has many local optima, they are close to one another in terms of the $l_2$-norm yielding similar values of MSE. This also implies that RFPI, which is guaranteed to find one of the local optima, performs nearly saturates as well (as measured by the MSE) as the signal recovery scheme based on (\ref{l1recovery}). 

\begin{figure}[t]
  \begin{center}
    \begin{tabular}{c}
    
      \begin{minipage}{0.5\hsize}
        \begin{center}
          \includegraphics[clip, angle=270,width=8.5cm]{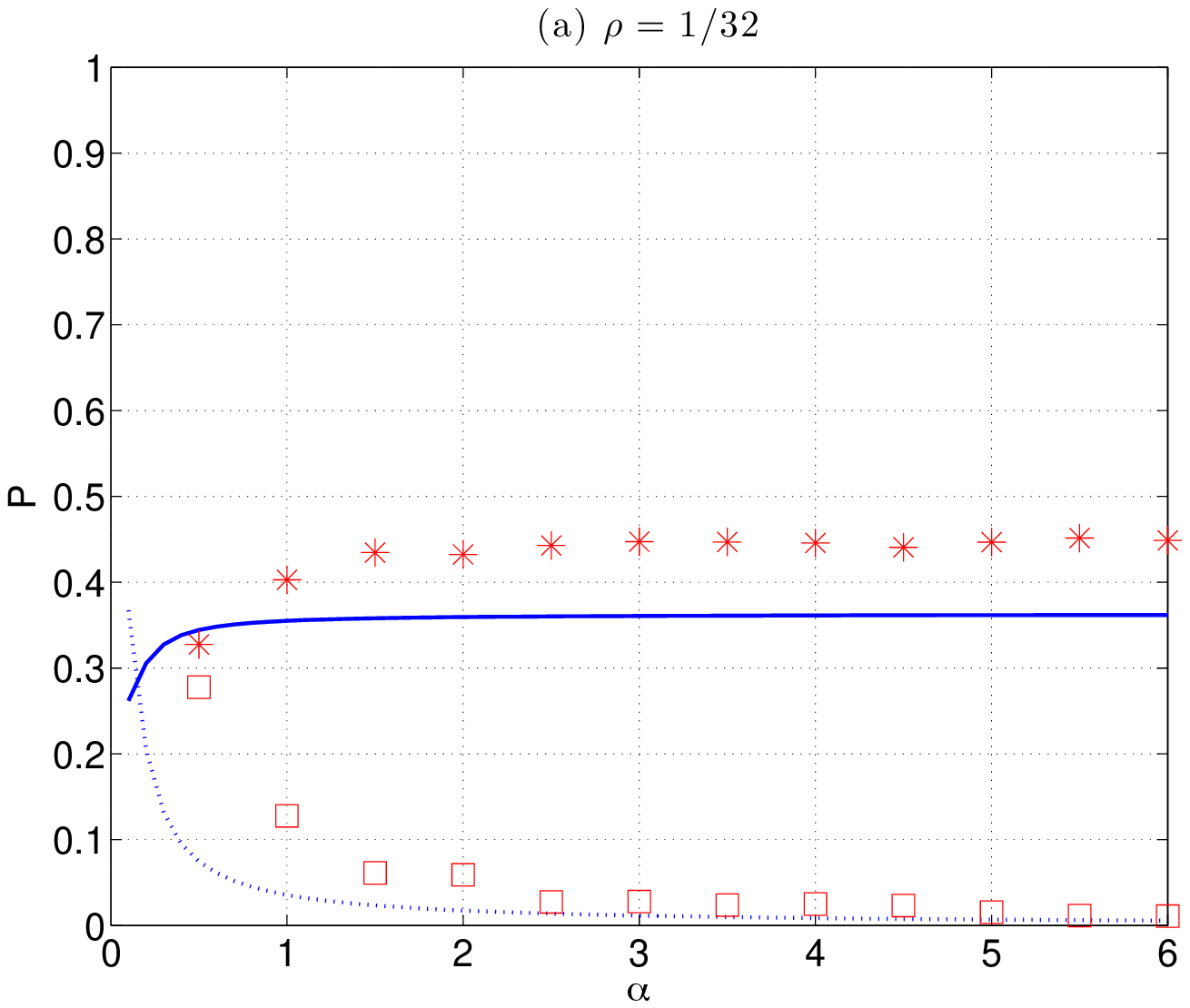}
          %\hspace{1.6cm} 
        \end{center}
      \end{minipage}
      
      \begin{minipage}{0.5\hsize}
        \begin{center}
          \includegraphics[clip, angle=270,width=8.5cm]{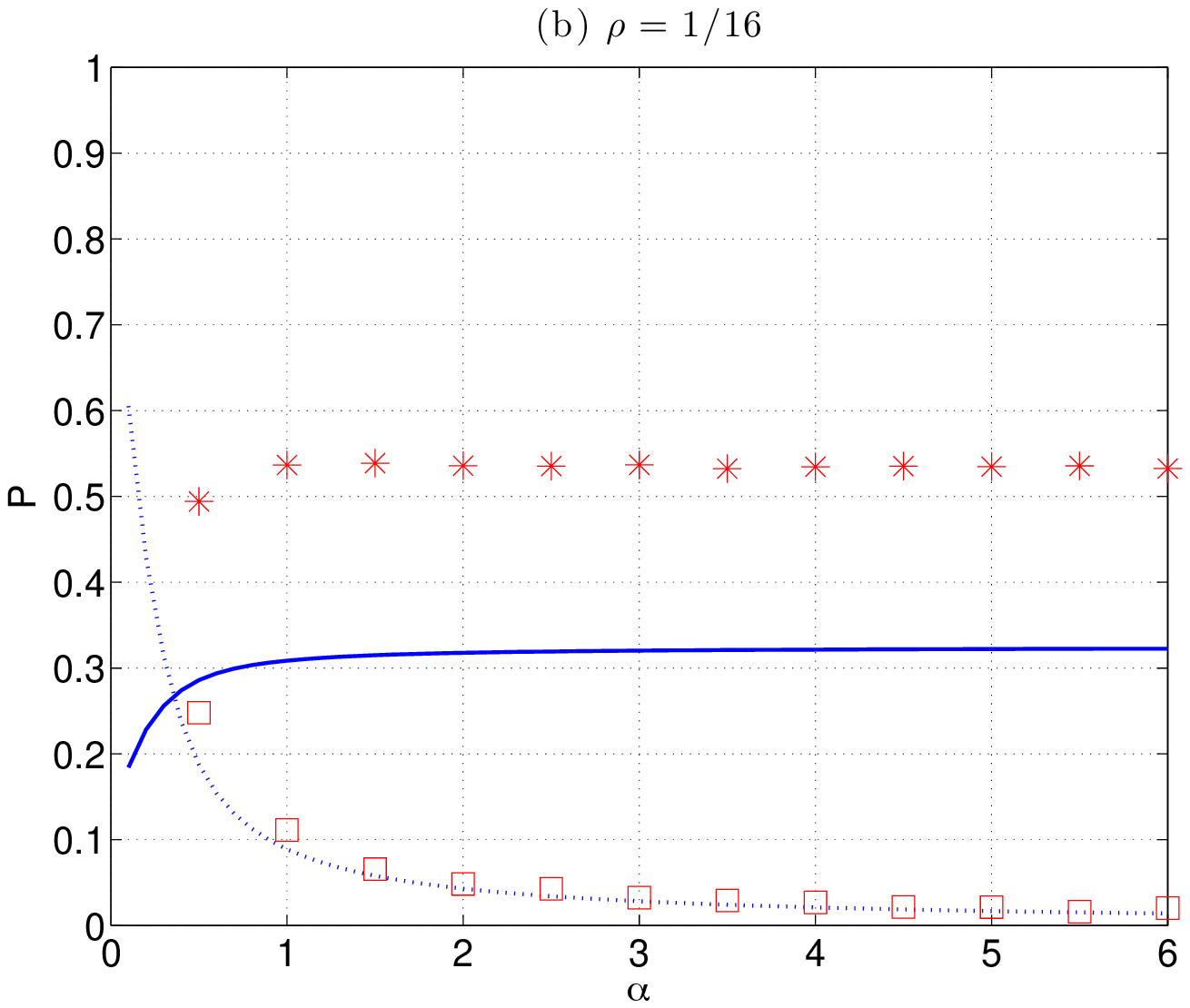}
          %\hspace{1.6cm}
        \end{center}
      \end{minipage}
\\
      \begin{minipage}{0.5\hsize}
        \begin{center}
          \includegraphics[clip, angle=270,width=8.5cm]{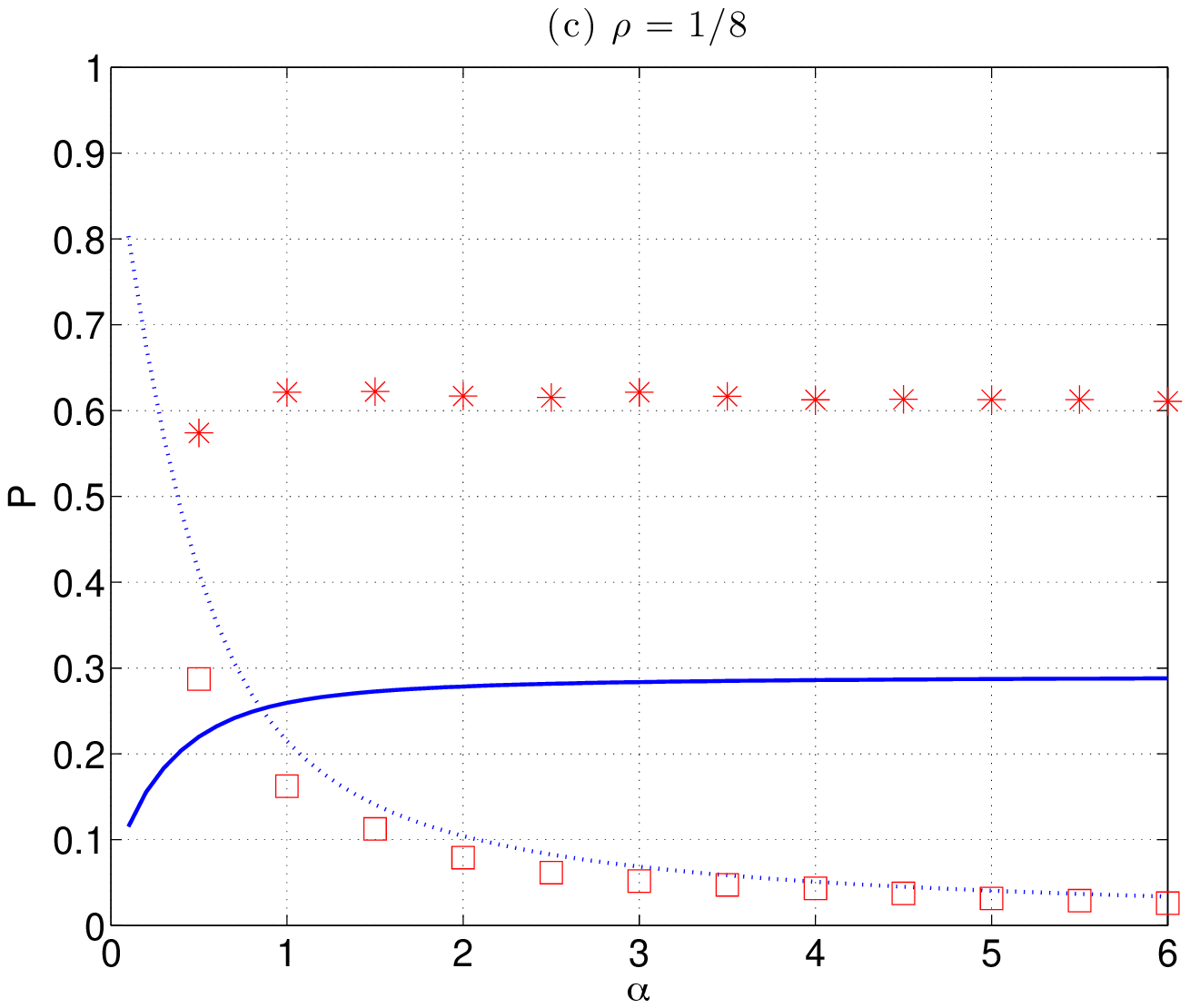}
          %\hspace{1.6cm} 
        \end{center}
      \end{minipage}
      
      \begin{minipage}{0.5\hsize}
        \begin{center}
          \includegraphics[clip, angle=270,width=8.5cm]{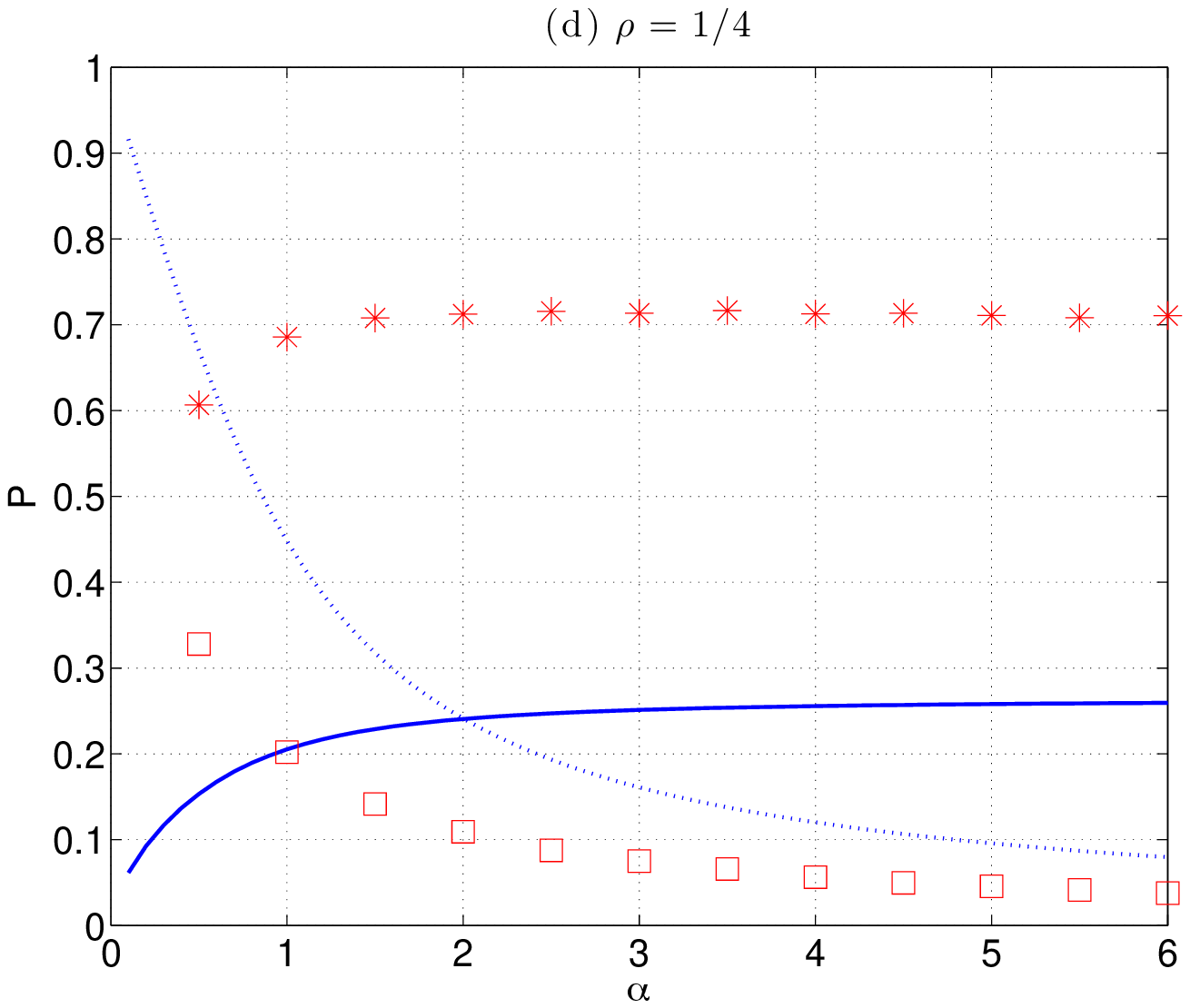}
          %\hspace{1.6cm} 
        \end{center}
      \end{minipage}
    \end{tabular}
\caption{\label{fig:L1ROC} FP and FN probabilities versus the measurement bit ratio $\alpha=M/N$. (a), (b), (c), and (d) corresponds to the 
$\rho=1/32, 1/16, 1/8$, and $1/4$ cases, respectively. Solid and dashed curves represent theoretical predictions obtained by the RS solution for FP and FN, respectively. Asterisks and squares denote experimental results for FP and FN, respectively. The experimental results were obtained by RFPI from 1000 samples for each condition of $N=128$ systems.}
\end{center}
\end{figure}

Of course, we have to keep in mind that the consistency between the theory and experiments depends highly on the performance measure used. Figures \ref{fig:L1ROC} (a)--(d) show the probabilities of wrongly predicting sites of nonzero and zero entries, which are sometimes referred to as false positive (FP) and false negative (FN), respectively. These indicate that there are considerably large discrepancies between the theory and experiments in terms of these performance measures, which is probably due to the influence of RSB. Nevertheless, the RS-based theoretical predictions are still qualitatively consistent with the experimental results in the way that the probability of a FP remains finite even when the measurement bit ratio $\alpha=M/N$ tends to infinity for any values of $\rho$. This implies that the $l_1$-based scheme is intrinsically unable to correctly identify sites of nonzero and zero entries. 

\section{Cavity-inspired signal recovery algorithm}
\label{CISR}
The analysis so far indicates that the performance of RFPI is good enough in the sense that there is little room for improvement in achievable MSE. RFPI requires tuning of two parameters $\delta$ and $\lambda$, however, which is rather laborious. In addition, the convergence of the inner loop of Figure \ref{RFPI} is relatively slow, which may limit its application range to systems of relatively small sizes. We therefore developed another recovery algorithm following the framework of the cavity method of statistical mechanics \cite{cavity,MezardMontanari2009}, or equivalently, the belief propagation of probabilistic inference \cite{BP3,BP1}. 

For simplicity of notations, let us first convert all the measurement results to $+1$ by multiplying $y_{\mu}$ $(\mu=1,2,\ldots, N)$ to each row of the measurement matrix $\vm{\Phi}=(\Phi_{\mu i})$ as $(\Phi_{\mu i}) \to (y_\mu \Phi_{\mu i})$, and newly denote the resultant matrix as $\vm{\Phi}=(\Phi_{\mu i})$.
In the new notation, introduction of Lagrange multipliers $\vm{a}=(a_\mu)$ and surplus variables $\vm{z}=(z_\mu)$ converts (\ref{l1recovery}) to an unconstrained optimization problem: 
\begin{eqnarray}
&&\mathop{\rm min}_{\vm{x},\vm{ z} > 0} \mathop{\rm max}_{\vm{a},\Lambda} 
\left \{
\sum_{i=1}^N |x_i|+\sum_{\mu=1}^M a_\mu \left (\sum_{i=1}^N \Phi_{\mu i} x_i-z_\mu \right )
+\frac{\Lambda}{2}\left (\sum_{i=1}^N x_i^2-N \right ) \right \} \cr
&& =
\mathop{\rm min}_{\vm{ x},\vm{ z} > 0} \mathop{\rm max}_{\vm{a},\Lambda} 
\left \{
\sum_{i=1}^N \left (\frac{\Lambda}{2} x_i^2 + |x_i| \right )-\sum_{\mu=1}^Ma_\mu z_{\mu}
+\sum_{\mu,i} \Phi_{\mu i} a_\mu x_i -\frac{N\Lambda}{2} \right \},   
\label{coupled_cost}
\end{eqnarray}
where $\vm{z}>0$ means that each entry of $\vm{z}$ is restricted to be positive. 

Coupling terms $\sum_{\mu i} \Phi_{\mu i} a_\mu x_i$ make the optimization of (\ref{coupled_cost}) a nontrivial problem. In statistical mechanics, a standard approach to resolving such a difficulty is to approximate 
(\ref{coupled_cost}) with a bunch of optimizations for single-body  
cost functions parameterized as 
\begin{eqnarray}
{\cal L}_i(x_i)=\frac{A_i}{2} x_i^2-H_i x_i +|x_i|, 
\label{x_cost}
\end{eqnarray}
and
\begin{eqnarray}
{\cal L}_\mu(a_\mu,z_\mu)
=-\frac{B_\mu}{2}a_\mu^2+K_\mu a_\mu-z_\mu a_\mu, 
\label{mu_cost}
\end{eqnarray}
where $A_i,B_\mu,H_i$, and $K_\mu$ are parameters to be determined in a self-consistent manner. 

In the cavity method this is done by introducing virtual systems that are defined by removing a single variable $x_i$ or a single pair of variables $(a_\mu, z_\mu)$ from the original system \cite{cavity}. When $N$ is sufficiently large, the law of large numbers allows us to assume that the values of $A_i$ and $B_\mu$ are constant independently of their indices; that is, that $A$ and $B$ are constants. Under this simplification, this method yields a set of self-consistent equations: 
\begin{eqnarray}
K_\mu &=& \sum_{i=1}^N \Phi_{\mu i} \hat{x}_i-B \hat{a}_\mu,  \label{BPnaive1} \\
\hat{a}_\mu &=& -\frac{1}{B} f^{\prime}\left (K_\mu   \right ),   \label{BPnaive2}\\
H_i &=& \sum_{\mu =1}^M \Phi_{\mu i} \hat{a}_\mu + \Gamma \hat{x}_i, \label{BPnaive3}\\
\hat{x}_i &=& \frac{1}{A} g^{\prime}\left (H_i  \right ),  \label{BPnaive4}
\end{eqnarray}
where $\mu=1,2,\ldots, M$, $i=1,2,\ldots,N$, $f(u)\equiv (u^2/2)\Theta(-u)$, and $g(u)\equiv \left ((|u|-1)^2/2 \right ) \Theta \left (|u|-1 \right )$. 
 $\Gamma$ is evaluated using $\{K_\mu\}$ and $B$ as
\begin{eqnarray}
\Gamma=B^{-1} \left (N^{-1} \sum_{\mu=1}^M f^{\prime \prime} \left (K_\mu \right ) \right ) =B^{-1} \left (N^{-1} \sum_{\mu=1}^M \Theta \left (|K_\mu | -1 \right ) \right ). \label{Gamma} 
\end{eqnarray}

$A$ is determined so that $\sum_{i=1}^N \hat{x}_i^2 =N$ holds in (\ref{BPnaive4}), 
 which provides 

$B$ as
\begin{eqnarray}
B=A^{-1}\left (N^{-1} \sum_{i=1}^N g^{\prime \prime} \left (H_i \right ) \right ) =A^{-1} \left (N^{-1} \sum_{i=1}^N \Theta \left (|H_i | -1\right ) \right ). \label{eq:B}
\end{eqnarray}

$\Gamma \hat{x}_i$ on the right-hand side of (\ref{BPnaive3}) is often referred to as the {\em Onsager reaction term} \cite{Onsager1,Onsager2}. Equation (\ref{BPnaive4}) offers the recovered signal. The derivations of these equations are provided in \ref{cavity_derivation}. 

A distinctive feature of the above set of equations is that they are free from tuning parameters such as $\lambda$ and $\delta$ in RFPI, which is highly beneficial in practical use. It is therefore unfortunate that in most cases the naive iterations of (\ref{BPnaive1})$\rightarrow$(\ref{BPnaive2}), (\ref{Gamma})$\rightarrow$(\ref{BPnaive3})$\rightarrow$(\ref{BPnaive4}), (\ref{eq:B})$\rightarrow$(\ref{BPnaive1})$\cdots$ hardly converge, which is considered a consequence of RSB \cite{Kabashima2003}, 
while a similar approach offers successful results for various other problems of compressed sensing
\cite{Rangan2011,Bayati2011}.

\begin{figure}
\renewcommand{\thepseudocode}{\arabic{pseudocode}}
\setcounter{pseudocode}{1}
\begin{pseudocode}[ruled]{Cavity-inspired Signal Recovery}{\textsc{B}, \mathbf{x}^*, \mathbf{H}^*}

1)\ \mbox{\bf Initialization}:\\
 \hspace{15pt}\text{X seed}: \hspace{65pt}\hat{\mathbf{x}}_{0} \GETS \hat{\mathbf{x}}^*\\
 \hspace{15pt}\text{H seed}: \hspace{65pt} \mathbf{H}_0 \GETS \mathbf{H}^* \\
\hspace{15pt}\text{Counter}: \hspace{62pt} k\GETS 0\\

2)\ \mbox{\bf Counter increase}:\\
 \hspace{30pt}k \GETS k+1\\

3)\ \mbox{\bf One-sided quadratic gradient descent}:\\

 \hspace{30pt}\mathbf{H}_{k}\GETS  \mathbf{H}_{k-1}-\textsc{B}^{-1}(\textsc{Y}\Phi)^{\rm T} 
 f^\prime\left (\textsc{Y}\Phi \hat{\mathbf{x}}_{k-1} \right ) \\

4)\ \mbox{\bf Assessment of Onsager coefficient}:\\
 \hspace{30pt}
 \Gamma \GETS (\textsc{NB})^{-1} \vm{1}^{\rm T} f^{\prime\prime} \left (\textsc{Y}\Phi \hat{\mathbf{x}}_{k-1} \right )\\

5)\ \mbox{\bf{Self-feedback \ cancellation}}:\\
 \hspace{30pt}\tilde{\mathbf{H}}_k \GETS \mathbf{H}_{k}+\Gamma \hat{\mathbf{x}}_{k-1}\\

6)\ \mbox{\bf Shrinkage ($l_1$-gradient descent)}:\\ 
 \hspace{30pt}(\mathbf{u})_i\GETS\text{sign}((\tilde{\mathbf{H}})_{i})
 \text{max}\{|(\tilde{\mathbf{H}})_i|-1,0\} \  \text{for all}\ i\\

7)\ \mbox{\bf Normalization}:\\
 \hspace{30pt}\hat{\mathbf{x}}_{k}\GETS \sqrt{N} \frac{\mathbf{u}}{||\mathbf{u}||_2}\\

8)\ \bold{Iteration}: \mbox{Repeat from 2) until convergence.}
\end{pseudocode}
\protect
\caption{\protect\label{proposedalgorithm}Pseudocode for the inner loop of the cavity-inspired signal recovery (CISR) algorithm. $\mathbf{x}^*$ and $\mathbf{H}^*$ are the convergent vectors of $\hat{\mathbf{x}}_k$ and $\hat{\mathbf{H}}_k$ obtained by the previous outer loop. The $\vm{1}$ in step 4) is the $N$-dimensional vector all entries of which are unity. If $(\mathbf{u})_i=0$ eventually holds for $\forall{i}$ in step 6), $\textsc{B}$ is reduced so that only $\mathop{\rm max}_{i} \{|(\mathbf{u})_i|\}$ becomes nonzero, and the procedure is restarted from step 3).}
\end{figure}

We found, however, that instead of updating $B$ by (\ref{eq:B}) at each iteration, handling $B$ as a parameter to be controlled in the outer loop, in conjunction with modifying (\ref{BPnaive1}) and (\ref{BPnaive2}) to 
\begin{eqnarray}
\hat{a}_\mu &=& \hat{a}_\mu -\frac{1}{B} f^{\prime}\left (\sum_{i=1}^N \Phi_{\mu i} \hat{x}_i  \right ),   \label{BPnaive2rev}
\end{eqnarray}
results in a fairly good approximate signal recovery algorithm. 

The necessity of controlling $B$ in the outer loop, which is essential for having good convergence in the inner loop, means that our algorithm still requires one tuning parameter. Nonetheless, the reduction in the number of the tuning parameters from two to one is considerably advantageous for practical use. In practice, the initial value of $B$ should be set so that only a single entry becomes nonzero. This is easily done by the Binary Iterative Hard Thresholding algorithm \cite{BIHT}, which requires the number of nonzero entries as extra prior knowledge. After the initial value is set, $B$ is reduced as $B_n=r B_{n-1}$ with an appropriate constant $0<r < 1$, where $n$ is the counter of the outer loop. The algorithm terminates when the difference between the convergent solutions of two successive outer loops is sufficiently small. 

The resultant algorithm is somewhat similar to RFPI as the combination of (\ref{BPnaive3}) and (\ref{BPnaive2rev}) roughly acts as the {\bf One-sided quadratic gradient descent} step in Figure \ref{RFPI}. However, as the length of $\vm{H}=(H_i)=\vm{\Phi}^{\rm T} \hat{\vm{a}}$ is not restricted to a fixed value, the current algorithm does not need a small step size $\delta$ for the convergence. Another significant difference from RFPI is the existence of the Onsager reaction term in (\ref{BPnaive3}). This term effectively cancels the self-feedback effects included in $H_i$ of (\ref{BPnaive3}), and this is expected to accelerate the convergence of the algorithm. A pseudocode for the inner loop is summarized in Figure \ref{proposedalgorithm}. 

\begin{figure}[t]
  \begin{center}
    \begin{tabular}{c}
    
      \begin{minipage}{0.5\hsize}
        \begin{center}
          \includegraphics[clip, angle=270,width=8.5cm]{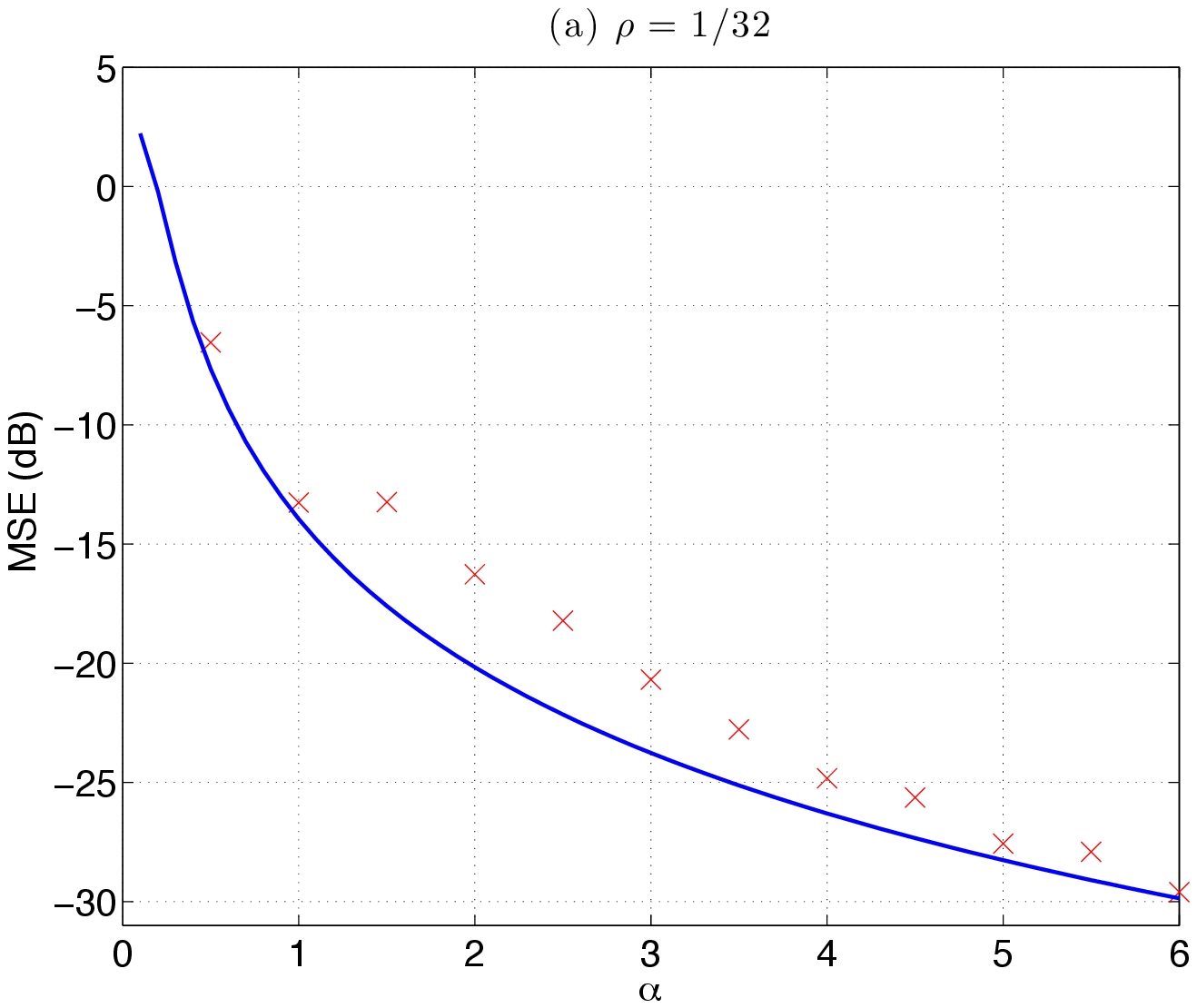}
          %\hspace{1.6cm} 
        \end{center}
      \end{minipage}
      
      \begin{minipage}{0.5\hsize}
        \begin{center}
          \includegraphics[clip, angle=270,width=8.5cm]{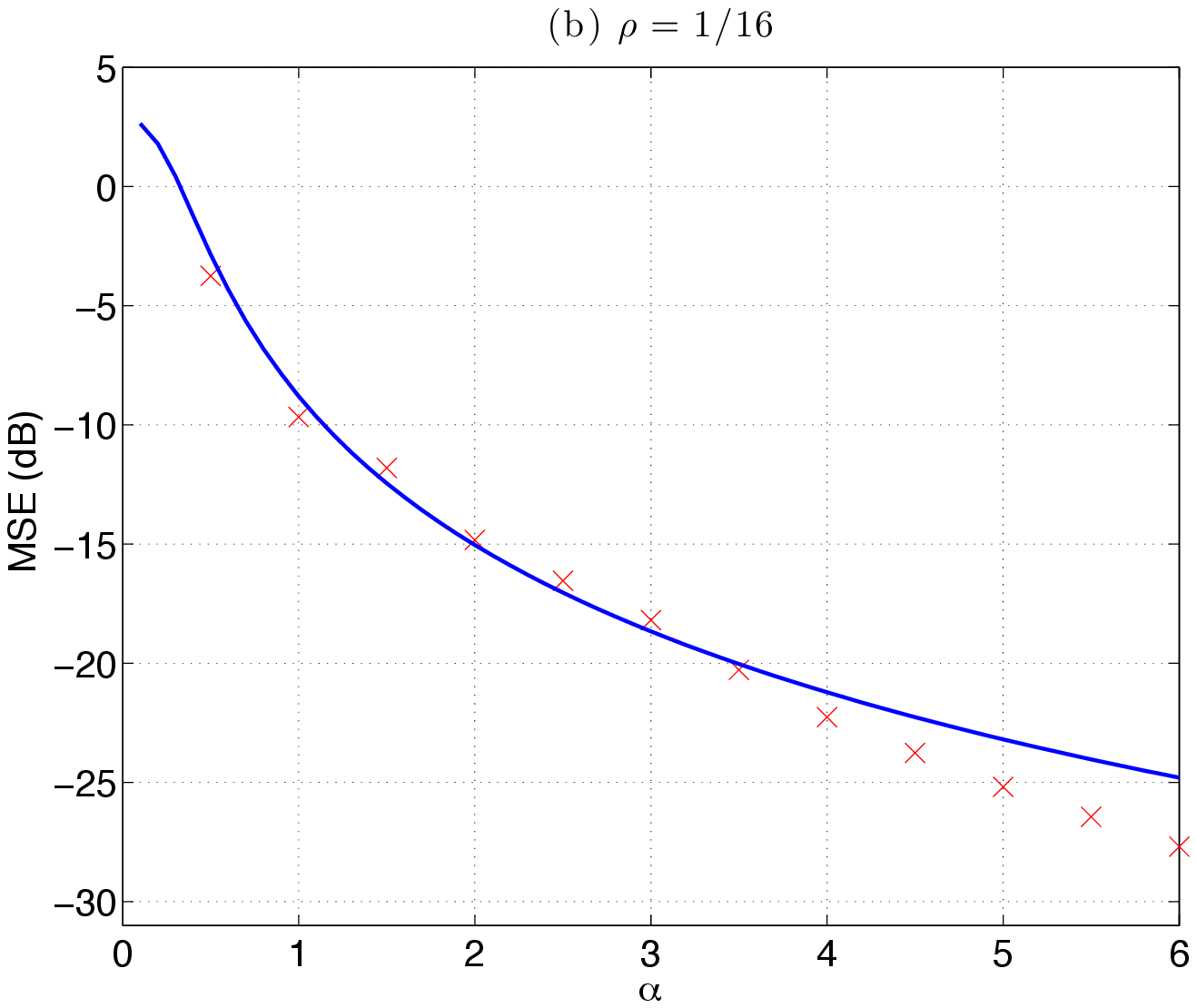}
          %\hspace{1.6cm}
        \end{center}
      \end{minipage} 
\\
      \begin{minipage}{0.5\hsize}
        \begin{center}
          \includegraphics[clip, angle=270,width=8.5cm]{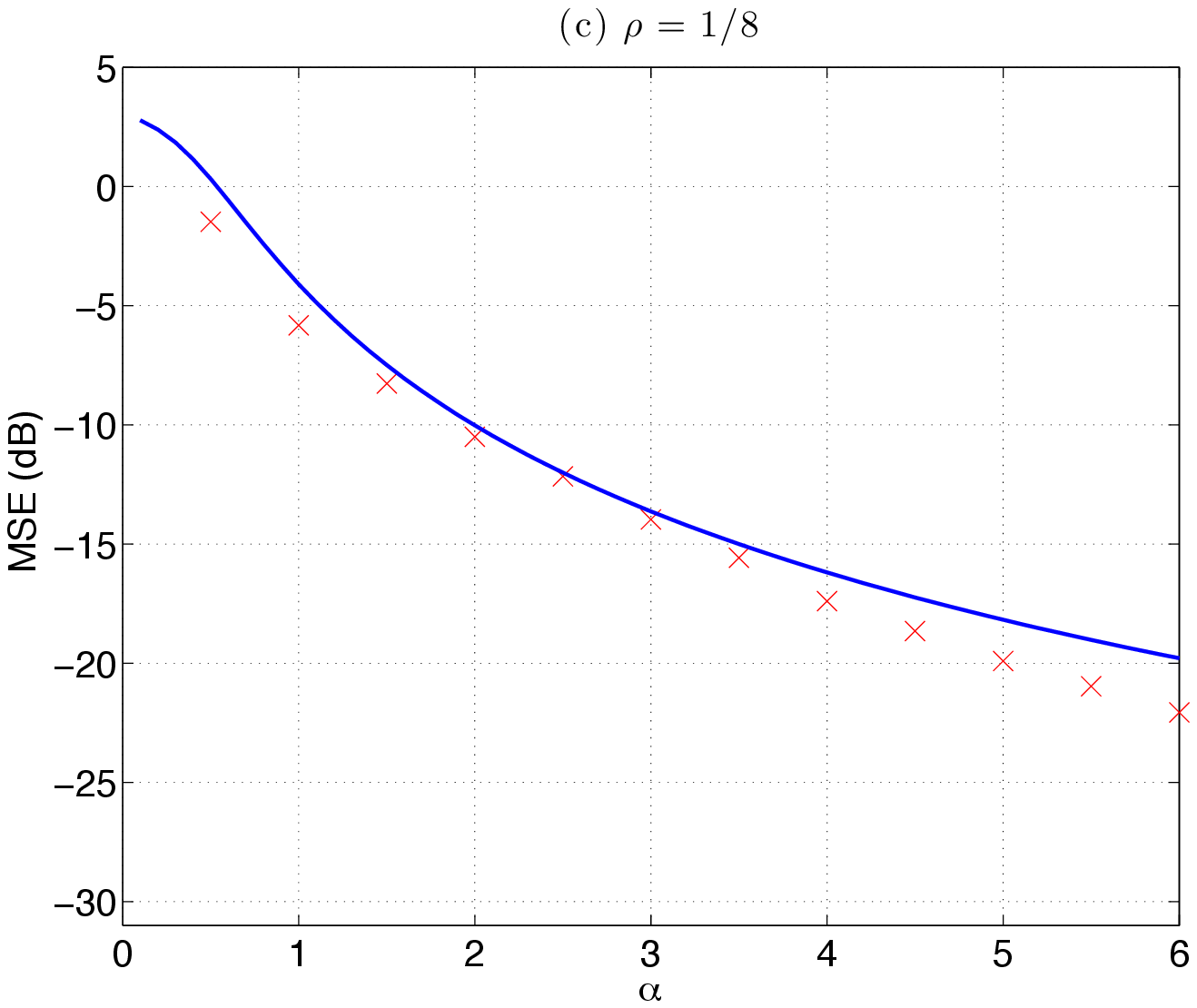}
          %\hspace{1.6cm} 
        \end{center}
      \end{minipage}
      
      \begin{minipage}{0.5\hsize}
        \begin{center}
          \includegraphics[clip, angle=270,width=8.5cm]{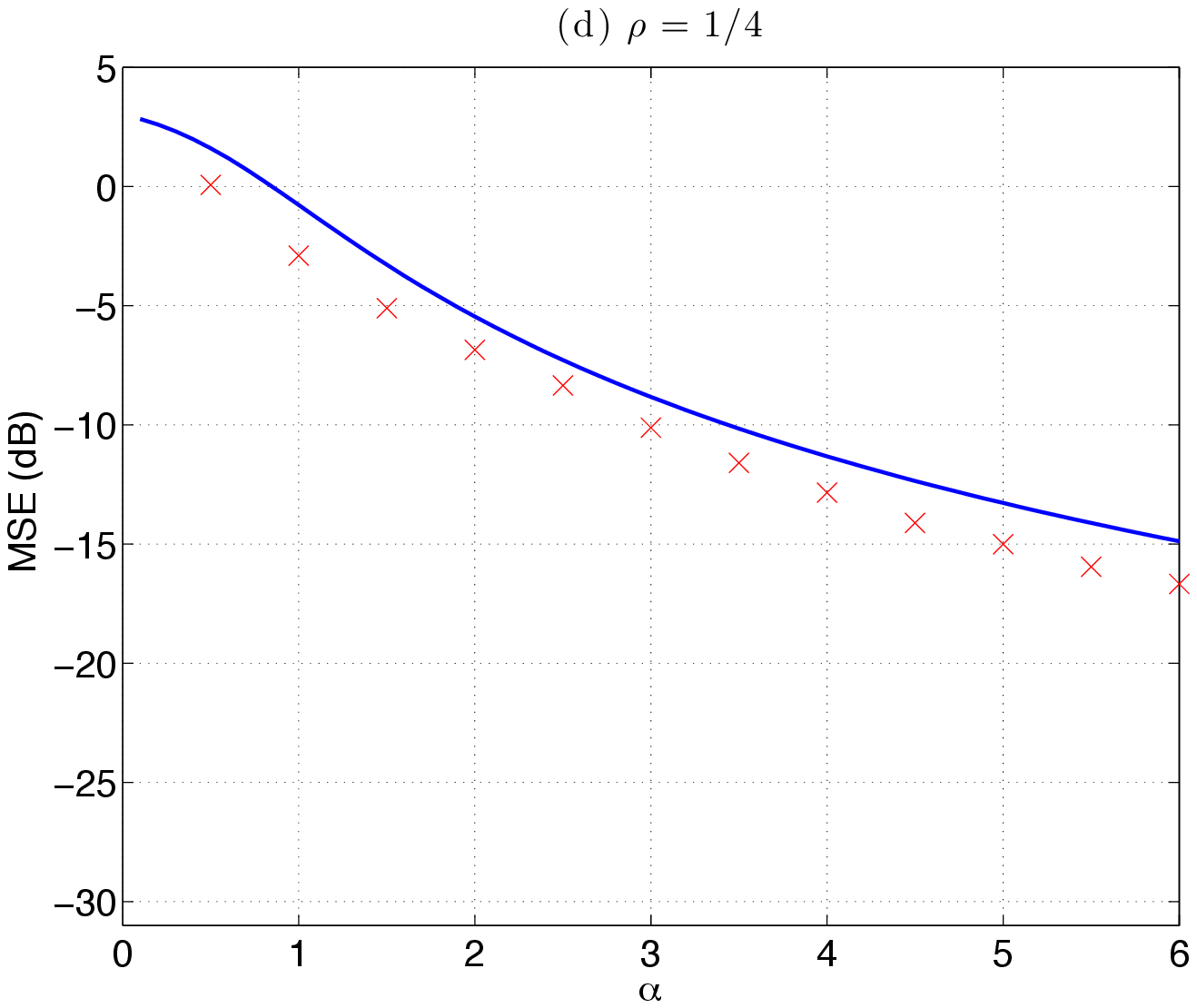}
          %\hspace{1.6cm} 
        \end{center}
      \end{minipage}
    \end{tabular}
\caption{\label{fig:BPMSE}MSE versus measurement bit ratio $\alpha$ for the cavity-inspired signal recovery (CISR) algorithm. Experimental conditions are the same as in Figures \ref{fig2} (a)--(d).} 
%[In the axis labels here,  MSE(dB)  -->  MSE (dB)]

  \end{center}
\end{figure}

\begin{figure}[t]
  \begin{center}
    \begin{tabular}{c}
    
      \begin{minipage}{0.5\hsize}
        \begin{center}
          \includegraphics[clip, angle=270,width=8.5cm]{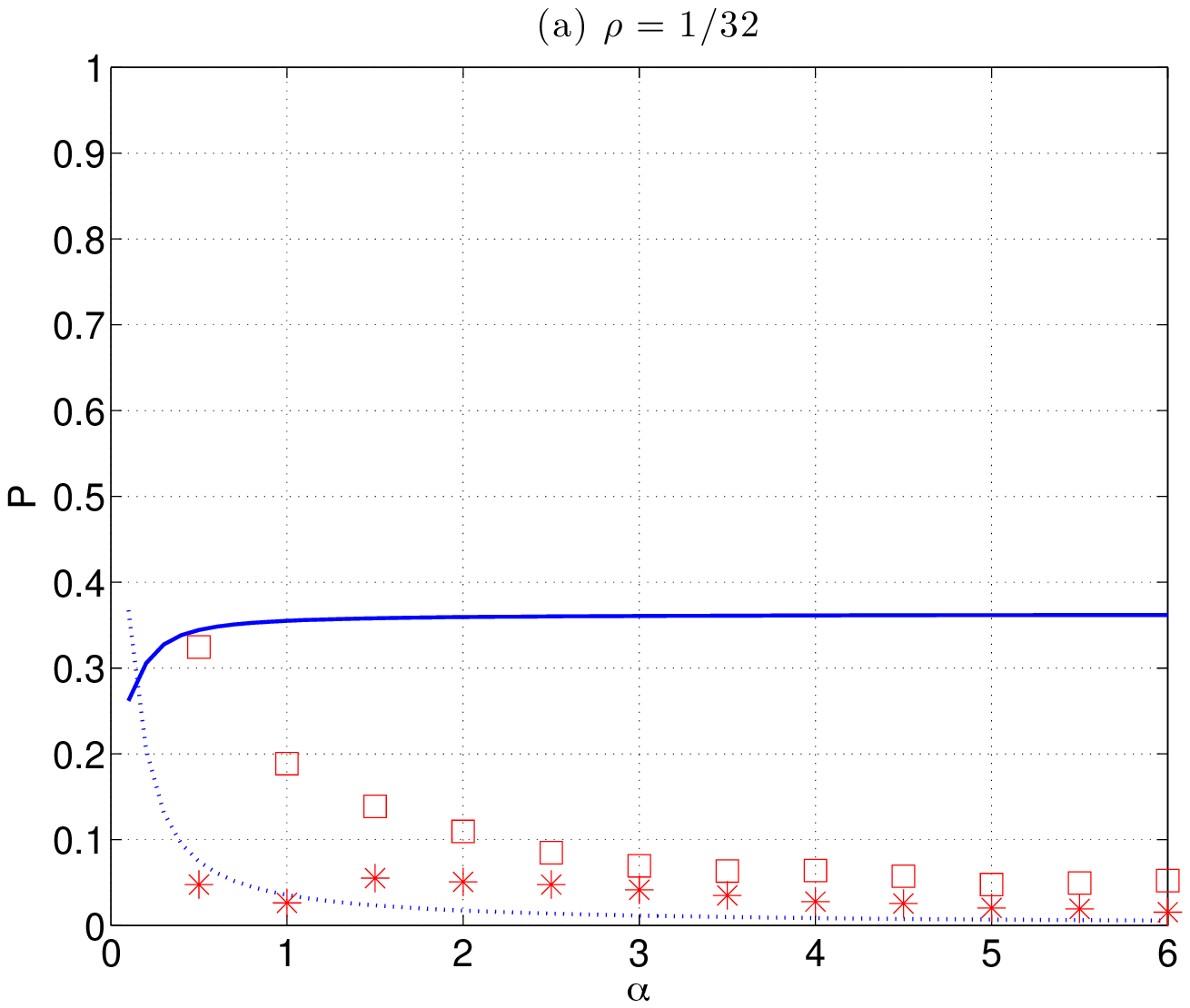}
          %\hspace{1.6cm} 
        \end{center}
      \end{minipage}
      
      \begin{minipage}{0.5\hsize}
        \begin{center}
          \includegraphics[clip, angle=270,width=8.5cm]{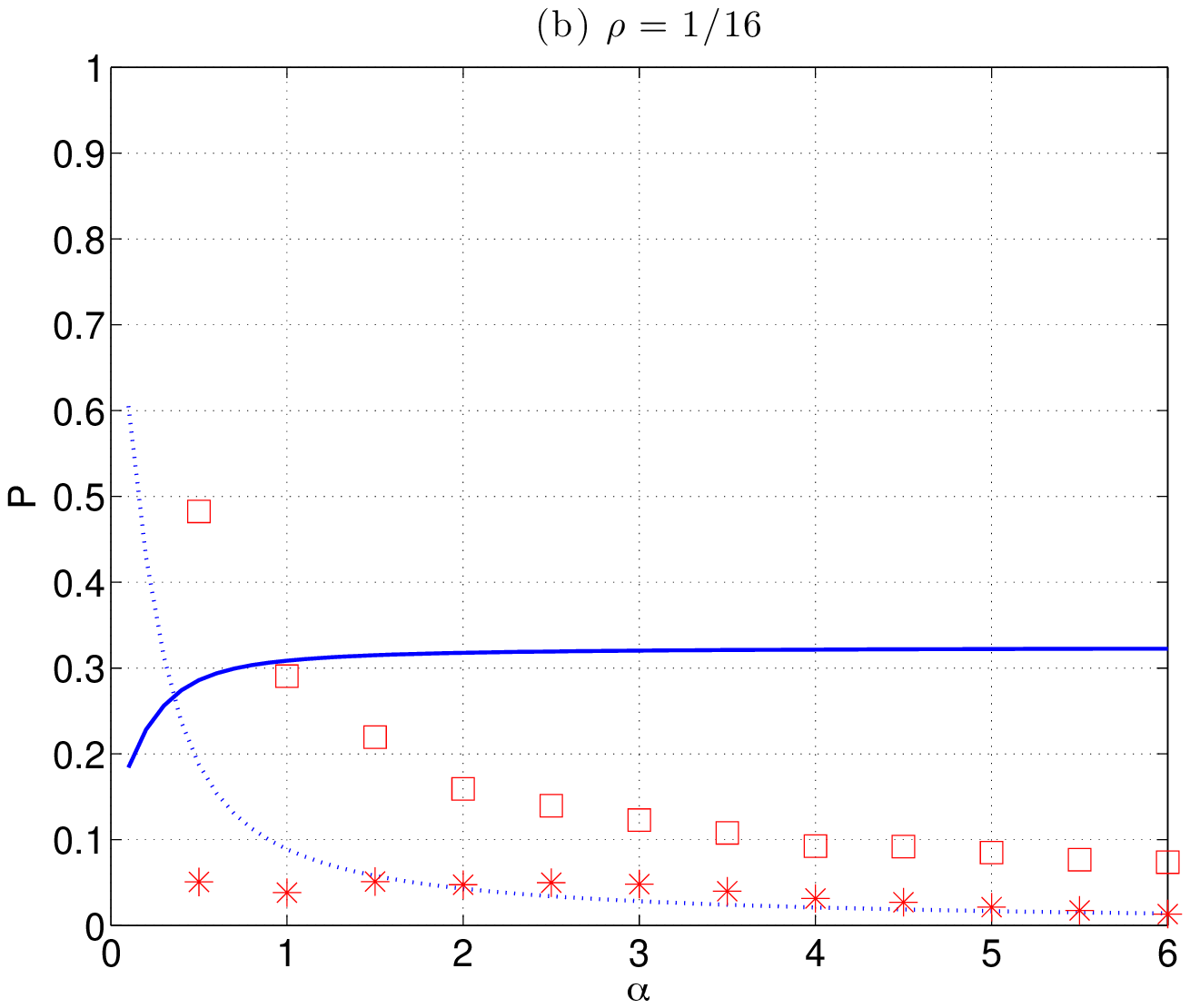}
          %\hspace{1.6cm}
        \end{center}
      \end{minipage}
\\
      \begin{minipage}{0.5\hsize}
        \begin{center}
          \includegraphics[clip, angle=270,width=8.5cm]{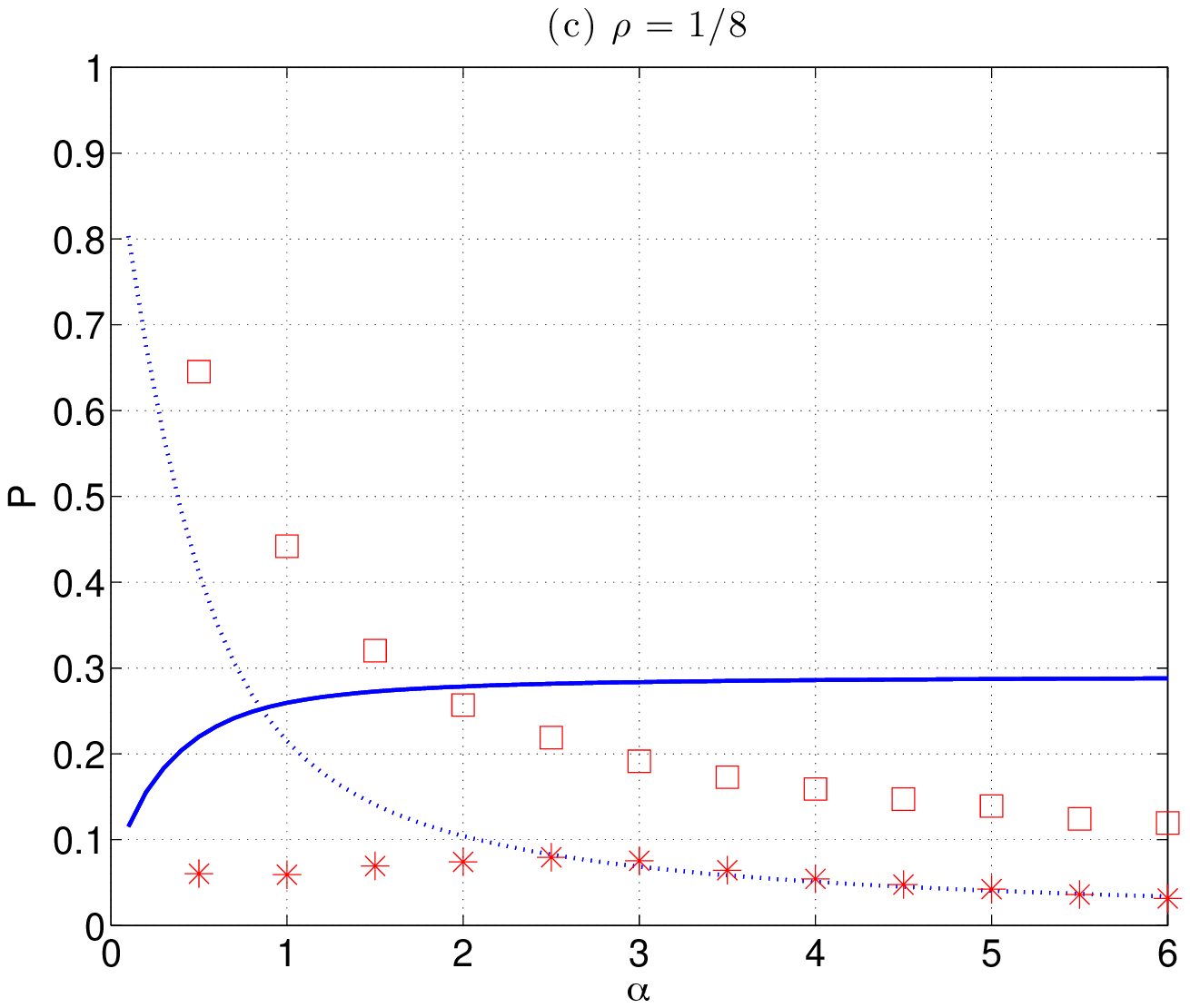}
          %\hspace{1.6cm} 
        \end{center}
      \end{minipage}
      
      \begin{minipage}{0.5\hsize}
        \begin{center}
          \includegraphics[clip, angle=270,width=8.5cm]{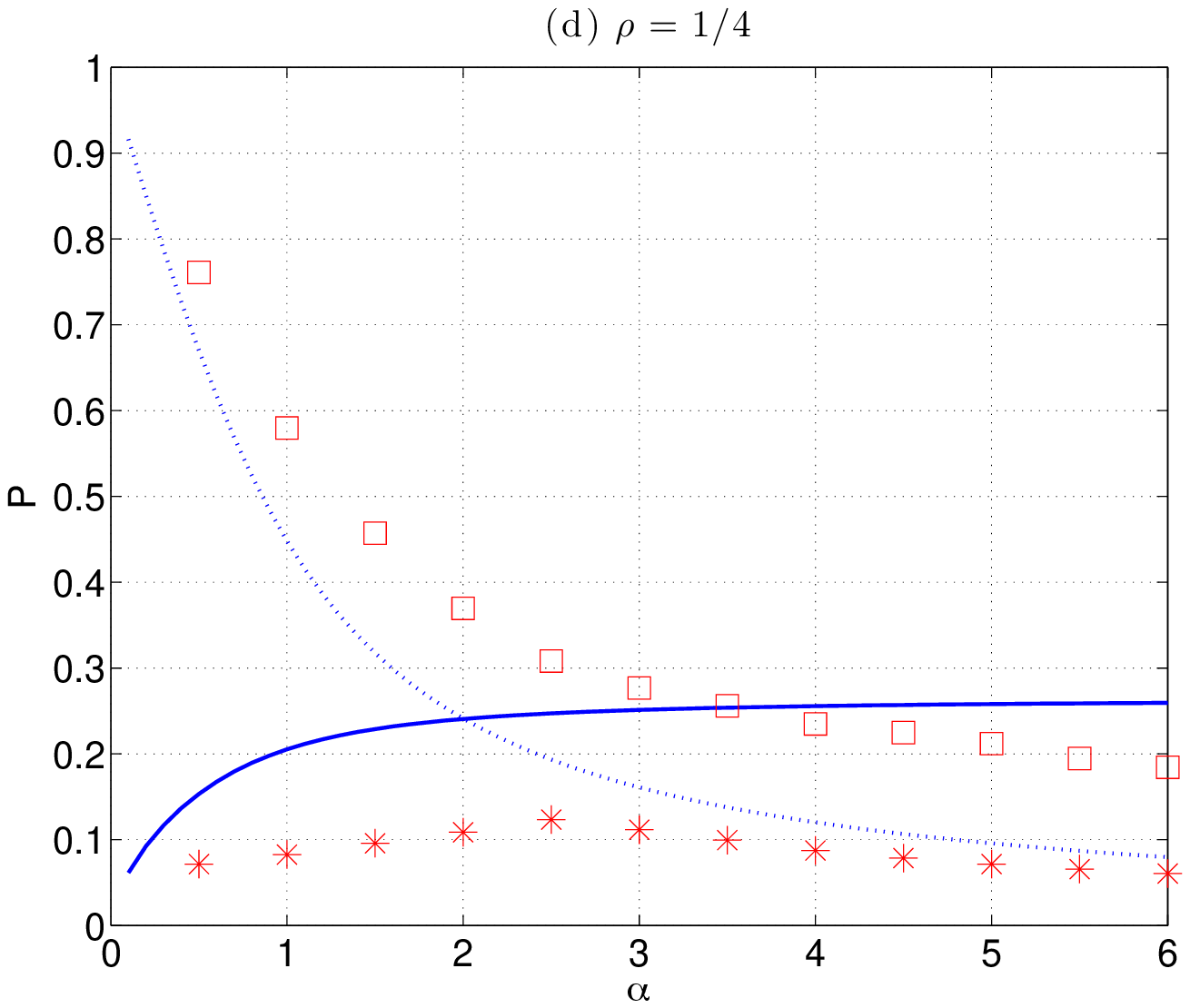}
          %\hspace{1.6cm} 
        \end{center}
      \end{minipage}
    \end{tabular}
\caption{\protect\label{fig:BPROC} FP and FN probabilities of versus measurement bit ratio $\alpha$ for the CISR.
Experimental conditions are the same as in Figures \ref{fig:L1ROC} (a)--(d).}
%[This caption seems incomplete. Won?t your target readers need as much information here as they did in the caption to Fig. 4?] 
\end{center}
\end{figure}

The MSE results obtained in numerical experiments with the cavity-inspired signal recovery (CISR) algorithm are shown in Figures \ref{fig:BPMSE} (a)--(d). They indicate that except in the case in which the nonzero density $\rho$ of the original signals is significantly low, 
CISR provides MSE values almost equal to or lower than those of RFPI. 
Figures \ref{fig:BPROC} (a)--(d) show the FP and FN probabilities for CISR. The discrepancies from the theoretical prediction are not unexpected because the modification of (\ref{BPnaive2}) to (\ref{BPnaive2rev}) means that CISR is no longer based on (\ref{l1recovery}) or (\ref{coupled_cost}). The FN probabilities for CISR are higher than those for RFPI, while the FP probabilities are lower. This implies that CISR has a capability of yielding sparser signals than RFPI, which is presumably because parameter $B$ of CISR is initially set so that only a single entry of $\hat{\vm{x}}$ is nonzero while such a tuning is not taken into account in RFPI. 
%This may also work for stopping the recovered signal at a closer point to the original one in terms of MSE when $\rho$ is relatively large. 
%[This is only arguably grammatical and is definitely unclear? too unclear for me to understand and recast into a clear statement consistent with the %data in Figs. 3 and 7. All I can do is hope that your target readers will be able to understand it anyway.] 

The run times actually required for performing the experiments in a MATLAB\textsuperscript{\textregistered} environment for the cases of $N=128$ and $M=3N=384$ are listed in Table \ref{table1}. Although the run times of RFPI may be reduced by optimally tuning the descent step size $\delta$, CISR is several hundreds of times faster than RFPI. This shows the significant computational efficiency of CISR. The NORT values in Table \ref{table1} are the run times when the Onsager reaction term in (\ref{BPnaive3}) was removed from CISR. Their being 1.13--2.37 times longer than those for CISR indicates that the cancellation of the self-feedback effects by adding the Onsager reaction term speeds the convergence of CISR significantly. 

\begin{table}
\protect
\caption{\protect\label{table1} Comparison of computational costs for the $N=128$ and $M=3N=384$ cases. The values listed here are the average run times (in seconds) evaluated in 1000 experiments, and the numbers in parentheses are the standard deviations. In RFPI, $\delta$ was roughly tuned as $0.01$, and $\lambda$ was enlarged as $\lambda_n=2 \lambda_{n-1}$ with the initial value $\lambda_0=0.005$ in the outer loop. On the other hand, $B$ of CISR was reduced as $B_n = 0.9 B_{n-1}$. The NORT values are %the for 
the run times required for performing the same experiments when the Onsager reaction term was removed from CISR. In all cases the algorithms terminated when the difference per entry, in terms of $l_1$-norm, between the convergent solutions of two successive outer loops was less than $10^{-8}$.}
\begin{center}
\begin{tabular}{|l|l|l|l|l|}
\hline 
& $K=4$ & $K=8$ & $K=16$ & $K=32$\\ \hline 
RFPI & $25.7636 (10.0799) $s & $27.8293 (3.3566)  $s & $33.3552 (3.2914) $s & $35.4574 (3.3869)$s \\
CISR & $0.0385 (0.0583) $s & $0.0705 (0.1058)  $s & $0.0245 (0.0346) $s & $0.0247 (0.0207)$ s\\
NORT & $0.0557 (0.0889) $s & $0.0795 (0.1095)  $s & $0.0581 (0.0566) $s & $0.0369 (0.0316) $s \\ 
\hline
\end{tabular}
\end{center}
\end{table}

\section{Summary}
In summary, we have examined typical properties of 1-bit compresses sensing (CS) proposed in \cite{1bitCS} utilizing methods of statistical mechanics. Signal recovery based on the $l_1$-norm minimization is a standard approach in CS research. Unlike the normal CS scheme, however, the $l_1$-based signal recovery cannot be formulated as a convex optimization problem, which makes practically performing it nontrivial. 

We have shown that the theoretical prediction of the performance of the $l_1$-based scheme, which is obtained by the replica method under the replica symmetric (RS) ansatz, exhibits a fairly good accordance (in terms of MSE) with experimental results obtained using for an approximate signal recovery algorithm, RFPI, proposed in \cite{1bitCS}. The replica symmetry of the RS solution turned out to be broken, however, which implies that there are many local optima for the optimization problem of the signal recovery. 
Our results suggest that the local optima, 
which can be searched by RFPI, yield similar values of MSE representing 
the potential performance limit of $l_1$-based recovery scheme.

We have also developed an approximate signal recovery algorithm utilizing the cavity method. Naive iterations of self-consistent equations derived directly from the cavity method hardly converge in most cases, which can be regarded as a consequence of the replica symmetry breaking. However, we have shown that modification of one equation in an appropriate manner, in conjunction with controlling a macroscopic variable in the outer loop, results in a fairly good signal recovery algorithm. Compared with RFPI, the resultant algorithm is beneficial in that the number of tuning parameters is reduced from two to one. Numerical experiments have also shown that whenever the density of nonzero entries of the original signal is not considerably small the cavity-inspired algorithm performs as well as or better than RFPI (in terms of MSE) and has a lower computational cost. 

We here focused on the $l_1$-based recovery scheme since it was proposed and examined in the seminal paper on 1-bit CS \cite{1bitCS}. However, the significance of the $l_1$-based scheme may be rather weak for 1-bit CS because the loss of convexity it entails keeps it from leading to the development of mathematically guaranteed and practically feasible algorithms. Therefore, much effort should be devoted to developing recovery algorithms following various principles. For example, the idea based on the Bayesian inference and matrix design that was proposed for standard CS \cite{Krzakala2012} may also be a promising approach for 1-bit CS. 

\ack
YX acknowledges a scholarship from Rotary Yoneyama Memorial Foundation, Inc. This study was partially supported by JSPS KAKENHI Nos. 22300003 and 22300098 (YK).
%%%%%%%%%%%%%%%%%%%%%%%%%%%%%%%appendix%%%%%%%%%%%%%%%%%%%%%%%%%%%%%%%%%%%%%%%%%%%%%
\appendix

\section{Derivation of $(\ref{eq:free energy})$}
\label{replicaderivation}
\subsection{Assessment of $\left[Z^{n}\left(\beta; \Phi, \textrm{\boldmath $x$}^{0}\right)\right]_{\Phi,\textrm{\boldmath $x$}^{0}}$ for $n \in \mathbb{N}$}
Averaging (\ref{eq:expansion}) with respect to $\vm{\Phi}$ and $\vm{x}^0$ offers the following expression of the $n$-th moment of the partition function:
\begin{eqnarray}
&&\left[Z^{n}\left(\beta; \Phi, \textrm{\boldmath $x$}^{0}\right)\right]_{\Phi,\textrm{\boldmath $x$}^{0}}=
\int %\prod_{a=1}^{n}
\prod_{a=1}^n \left (d \textrm{\boldmath $x$}^a\delta\left(\vert \textrm{\boldmath $x$}^a\vert^{2}-N\right) \times 
e^{-\beta||\textrm{\boldmath $x^a$}||_{1}} \right ) \cr
&& \hspace*{3cm} \times 
\left[\prod_{a=1}^n \prod_{\mu=1}^M 
\Theta\left(
(\vm{\Phi} \vm{x}^0)_\mu 
(\vm{\Phi} \vm{x}^a)_\mu 
\right)
\right]_{\vm{\Phi},\vm{x}^{0}}. 
\label{Zmoment}
\end{eqnarray}
We insert $n(n+1)/2$ trivial identities
\begin{eqnarray}
1=N \int dq_{ab} \delta \left (\vm{x}^a \cdot \vm{x}^b - N q_{ab} \right ), 
\end{eqnarray}
where $a>b=0,1,2,\ldots,n$, into (\ref{Zmoment}). Furthermore, we define a joint distribution of $n+1$ vectors $\{\vm{x}^a\}=\{\vm{x}^0, \vm{x}^1,\vm{x}^2,\ldots,\vm{x}^n \}$ as
\begin{eqnarray}
&&P\left (\{\vm{x}^a\} |\vm{Q}\right )= \frac{1}{V\left (\vm{Q}\right ) } P(\vm{x}^0) \times 
\prod_{a=1}^n \left (\delta \left (\left |\vm{x}^a\right |^2 -N \right ) 
\times 
e^{-\beta||\textrm{\boldmath $x^a$}||_{1}} \right )  \cr
&& \hspace*{3cm} \times \prod_{a>b} \delta \left (\vm{x}^a \cdot \vm{x}^b - N q_{ab} \right ), 
\label{replica_joint_dist}
\end{eqnarray}
where $\vm{Q}=(q_{ab})$ is an $(n+1) \times (n+1)$ symmetric matrix whose $00$ and the other diagonal entries are fixed as $\rho$ and $1$, respectively. $P(\vm{x}^0)=\prod_{i=1}^N \left ((1-\rho)\delta(x_i^0)+\rho \tilde{P}(x_i^0) \right )$ denotes the distribution of the original signal $\vm{x}^0$, and $V\left (\vm{Q}\right )$ is the normalization constant that makes $\int \prod_{a=0}^n d \vm{x}^a P\left (\{\vm{x}^a\} |\vm{Q}\right )=1$ hold. These indicate that (\ref{Zmoment}) can also be expressed as
\begin{eqnarray}
\left[Z^{n}\left(\beta; \Phi, \textrm{\boldmath $x$}^{0}\right)\right]_{\Phi,\textrm{\boldmath $x$}^{0}}
=\int d\vm{Q} \left (V\left (\vm{Q}\right ) \times \Xi\left (\vm{Q}\right ) \right ), 
\label{Zn}
\end{eqnarray}
where $d\vm{Q} \equiv \prod_{a>b}d q_{ab}$ and 
\begin{eqnarray}
\Xi\left (\vm{Q}\right )
=\int \prod_{a=0}^n d\vm{x}^a P\left (\{\vm{x}^a\} |\vm{Q} \right ) 
\left[\prod_{a=1}^n \prod_{\mu=1}^M 
\Theta\left(
(\vm{\Phi} \vm{x}^0)_\mu 
(\vm{\Phi} \vm{x}^a)_\mu 
\right)
\right]_{\vm{\Phi}}. 
\label{Xi}
\end{eqnarray}

Equation (\ref{Xi}) can be regarded as the average of $\prod_{a=1}^n \prod_{\mu=1}^M \Theta\left((\vm{\Phi} \vm{x}^0)_\mu (\vm{\Phi} \vm{x}^a)_\mu \right)$ with respect to $\{\vm{x}^a\}$ and $\vm{\Phi}$ over distributions of $P\left (\{\vm{x}^a\} \right )$ and $P(\vm{\Phi}) \equiv \left (\sqrt{2\pi/N} \right )^{-MN} \exp \left (-(N/2) \sum_{\mu,i} \Phi_{\mu i}^2 \right )$. In computing this, it is noteworthy that the central limit theorem guarantees that $u_\mu^a \equiv (\vm{\Phi} \vm{x}^a)_\mu = \sum_{i=1}^N \Phi_{\mu i} x_i^a$ can be handled as zero-mean multivariate Gaussian random numbers whose variance and covariance are provided by 
\begin{eqnarray}
\left [u_\mu^a u_\nu^b \right ]_{\vm{\Phi},\{\vm{x}^a\}}
=\delta_{\mu \nu} q_{ab}, 
\end{eqnarray}
when $\vm{\Phi}$ and $\{\vm{x}^a\}$ are generated independently from $P(\vm{\Phi})$ and $P\left (\{\vm{x}^a\} \right )$, respectively. This means that (\ref{Xi}) can be evaluated as
\begin{eqnarray}
\Xi(\vm{Q}) &=&\left (\frac{\int d \vm{u} \exp\left (-\frac{1}{2}\vm{u}^{\rm T} \vm{Q}^{-1} \vm{u} \right )
\prod_{a=1}^n \Theta\left (u^0 u^a \right )  }{
(2\pi)^{(n+1)/2} (\det \vm{Q})^{1/2}}
\right )^M \cr
&=& \left (2 \int \frac{ d \vm{u}\exp\left (-\frac{1}{2}\vm{u}^{\rm T} \vm{Q}^{-1} \vm{u} \right )
\Theta\left (u^0 \right )\prod_{a=1}^n \Theta\left (u^a \right )  }{
(2\pi)^{(n+1)/2} (\det \vm{Q})^{1/2}}
 \right )^M. 
\label{logXi}
\end{eqnarray}

On the other hand, expressions 
\begin{eqnarray}
\delta\left (|\vm{x}^a|^2-N \right )
=\frac{1}{4 \pi} \int_{-{\rm i}\infty}^{+{\rm i}\infty}
d \hat{q}_{aa} \exp \left (-\frac{1}{2} \hat{q}_{aa} 
\left (|\vm{x}^a|^2-N \right ) \right )
\end{eqnarray}
and 
\begin{eqnarray}
\delta\left (\vm{x}^a \cdot \vm{x}^b -N q_{ab}\right )
=\frac{1}{2 \pi} \int_{-{\rm i}\infty}^{+{\rm i}\infty}
d \hat{q}_{ab} \exp \left ( \hat{q}_{ab} 
\left (\vm{x}^a \cdot \vm{x}^b-N q_{ab} \right ) \right ), 
\end{eqnarray}
and use of the saddle point method offer 
\begin{eqnarray}
&&\frac{1}{N} \ln V(\vm{Q})
=\mathop{\rm extr}_{\hat{\vm{Q}}}
\left \{
-\frac{1}{2} {\rm Tr} \hat{\vm{Q}}\vm{Q} \right . \cr
&& \hspace*{2cm}
\left . + \ln 
\left (
\int d\vm{x} 
P(x^0) \exp \left (\frac{1}{2} 
\vm{x}^{\rm T} \hat{\vm{Q} }\vm{x}-\beta \sum_{a=1}^n \beta |x^a| \right )
\right ) 
\right \}.  
\label{logV}
\end{eqnarray}
Here $\vm{x}=(x^0,x^1,\ldots,x^n)^{\rm T}$ and $\hat{\vm{Q}}$ is an $(n+1)\times (n+1)$ symmetric matrix whose $00$ and other diagonal components are given as $0$ and $-\hat{q}_{aa}$, respectively, while off-diagonal entries are offered as $\hat{q}_{ab}$. Equations (\ref{logXi}) and (\ref{logV}) indicate that $N^{-1} \ln \left [Z^n (\beta;\vm{\Phi}, \vm{x}^0 ) \right]_{\vm{\Phi}, \vm{x}^0}$ is correctly evaluated by using the saddle point method with respect to $\vm{Q}$ in the assessment of the right-hand side of (\ref{Zn}) when $N$ and $M$ tend to infinity keeping $\alpha=M/N$ finite. 

\subsection{Treatment under the replica symmetric ansatz}
Let us assume that the relevant saddle point in assessing (\ref{Zn}) is of the form of (\ref{RSanzats}) and, accordingly, 
\begin{eqnarray}
\hat{q}_{ab}=\hat{q}_{ba} =\left \{
\begin{array}{ll}
0, &( \mbox{$a=b=0$})\\
\hat{m}, &( \mbox{$a=1,2,\ldots,n$; $b=0$})\\
\hat{Q}, &( \mbox{$a=b=1,2,\ldots,n$})\\
\hat{q}, &( \mbox{$a\ne b =1,2,\ldots,n$})
\end{array}
\right .  .
\label{RShatQ}
\end{eqnarray}
$n+1$ dimensional Gaussian random variables $u^0,u^1,\ldots u^n$ whose variance and covariance are provided as (\ref{RSanzats}) can be expressed as 
\begin{eqnarray}
&& u^0=\sqrt{\rho-\frac{m^2}{q}}s^0 + \frac{m}{\sqrt{q}} z, \label{Newgauss0}\\
&& u^a=\sqrt{1-q} s^a+ \sqrt{q} z, \ (a=1,2,\ldots,n) \label{Newgaussa} 
\end{eqnarray}
utilizing $n+2$ independent standard Gaussian random variables $z$ and $s^0,s^1,\ldots,s^n$. This indicates that (\ref{logXi}) is evaluated as
\begin{eqnarray}
\Xi(\vm{Q})=\left (
2 \int {\rm D}z 
H\left (\frac{m}{\sqrt{\rho q-m^2} }z \right )
H^n \left (\sqrt{\frac{q}{1-q}} z \right ) \right )^M. 
\label{NewXi}
\end{eqnarray}
On the other hand, substituting (\ref{RShatQ}) into (\ref{logV}), in conjunction with the identity
\begin{eqnarray}
\exp \left (\hat{q} \sum_{a>b(\ge 1)} x^a x^b \right )=\int {\rm D}z \exp \left (\sum_{a=1}^n 
\left (
-\frac{\hat{q}}{2} (x^a)^2 + \sqrt{\hat{q}} z x^a \right ) \right ) 
\end{eqnarray}
provides 
\begin{eqnarray}
&&\frac{1}{N} \ln V(\vm{Q} )=\mathop{\rm extr}_{\hat{Q}, \hat{q},\hat{m} }
\left \{
\frac{n}{2}\hat{Q}-\frac{n(n-1)}{2} \hat{q}q -\hat{m} m \right . \cr
&&\left . 
+\ln 
\left [\left (
\!
\int 
\!
dx \exp 
\!\left (
\!
-\frac{\hat{Q}
\!+\! \hat{q}}{2} x^2\!+\!\left (\!\sqrt{\hat{q}}z \!+\! \hat{m} x^0 \!\right )
\!x \!-\!\beta |x| \right )
\right )^n \right ]_{x^0,z} \right \}. 
\label{NewV}
\end{eqnarray}
Although we have assumed that $n \in \mathbb{N}$, the expressions of (\ref{NewXi}) and (\ref{NewV}) are likely to hold for $n \in \mathbb{R}$ as well. Therefore the average free energy $\overline{f}$ can be evaluated by substituting these expressions into the formula $\overline{f}= -\lim_{n\to 0} (\partial/\partial n) \left ((\beta N)^{-1} \ln \left [Z^n (\beta; \vm{\Phi},\vm{x}^0 ) \right]_{\vm{\Phi},\vm{x}^0} \right )$. 

In the limit of $\beta \to \infty$, a nontrivial saddle point is obtained only when $\chi \equiv \beta (1-q)$ is kept finite. Accordingly, we change the notations of the auxiliary variables as $\hat{Q}+\hat{q} \to \beta \hat{Q}$, $\hat{q} \to \beta^2 \hat{q}$, and $\hat{m} \to \beta \hat{m}$. Furthermore, we use the asymptotic forms
\begin{eqnarray}
&&\lim_{\beta \to \infty}\frac{1}{\beta}\int {\rm D}z
H\left (\frac{m}{\sqrt{\rho q-m^2} }z \right )
\ln H\left (\sqrt{\frac{q}{1-q}} z \right ) \cr
&&=\int {\rm D}z
H\left (\frac{m}{\sqrt{\rho -m^2} }z \right )\left (-\frac{z^2}{2\chi} \Theta(z) \right ) \cr
&&= -\frac{1}{4 \pi \chi}\left (\arctan \left (\frac{\sqrt{\rho-m^2}}{m} \right )
-\frac{m}{\rho}\sqrt{\rho-m^2} \right ) 
\end{eqnarray}
and 
\begin{eqnarray}
&&\lim_{\beta \to \infty}\frac{1}{\beta}
\ln 
\left (
\!
\int 
\!
dx \exp 
\!\left (\beta \left (
\!
-\frac{\hat{Q}}{2} x^2\!+\!\left (\!\sqrt{\hat{q}}z \!+\! \hat{m} x^0 \!\right )
\!x \!-\! |x| \right ) \right )
\right ) \cr
&& =-\phi\left (\sqrt{\hat{q}}z+\hat{m}x^0;\hat{Q} \right ). 
\end{eqnarray}
Using these in the resultant expression of $\overline{f}$ offers (\ref{eq:free energy}). 

\section{Stability of the RS solution}
\label{RSstability}
The 1-step replica symmetry breaking (1RSB) ansatz means that, at the relevant saddle point, $n$ replica indices $1,2,\ldots,n$ are classified into $n/p$ groups of an equal size $p$, and $q_{ab}=q_1$ holds if $a$ and $b$ belong to an identical group and $q_0(\le q_1)$, otherwise. This yields the following expression of the average free energy of finite temperature:
\begin{eqnarray}
&&\overline{f}=\mathop{\rm extr}_{\omega}\left \{
-\frac{1}{\beta  } \left [ \ln \left (\int {\rm D}{t}
\exp \left (-p {\cal Y}_0 \right )
\right ) \right ]_{x^0,z} \right  . \cr
&& 
\hspace*{0cm}-\frac{1}{2\beta}(\hat{Q}+\hat{q}_1)+\frac{\hat{q}_1}{2\beta} (1-q_1)+\frac{p}{2\beta}
(\hat{q}_1q_1-\hat{q}_0q_0 )+\frac{1}{\beta}\hat{m}m\cr
&&
\hspace*{0cm}\left . 
-\frac{2\alpha}{\beta p}\!
\int \! {\rm D}z H\left (\!\frac{m }{\sqrt{\rho q-m^2}}z\! \right )
\ln \left (\int {\rm D}t \exp \left (-p {\cal Y}_1 \right ) \right )
\right \}, 
\label{1RSB}
\end{eqnarray}
where ${\cal Y}_0\equiv -\ln \left (\int dx \exp \left (-(\hat{Q}+\hat{q}_1)x^2 /2 + (\sqrt{\hat{q}_1-\hat{q}_0}t+\sqrt{\hat{q}_0}z +\hat{m}x^0 )x-\beta |x| \right ) \right )$, ${\cal Y}_1\equiv - \ln \left (\int {\rm D}x \Theta\left (-\left (\sqrt{1-q_1}x+\sqrt{q_1-q_0}t+\sqrt{q}_0 z \right ) \right ) \right )$, $\omega=\{q_1,q_0,m,\hat{Q},\hat{q}_1, \hat{q}_0, \hat{m} \}$, and $\left [\cdots \right]_{x^0,z}=\int dx^0 P(x^0)\int {\rm D}z \left (\cdots \right )$. The RS solution is regarded as a special case of the 1RSB solution for which $q_1=q_0 $ holds. Therefore one can check the thermodynamical validity of the RS solution by examining the stability of the solution of $q_1=q_0$ under the 1RSB ansatz. 

The extremization condition of (\ref{1RSB}) indicates that 
\begin{eqnarray}
&&q_1-q_0=\left [ \frac{\int {\rm D}t e^{-p {\cal Y}_0}
\left (\partial{\cal Y}_0/\partial (\sqrt{\hat{q}_0} z) \right )^2}
{\int {\rm D}t e^{-p {\cal Y}_0}} \right . \cr
&&\hspace*{4cm} \left . -
\left (\frac{\int {\rm D}t e^{-p {\cal Y}_0}
\left (\partial{\cal Y}_0/\partial (\sqrt{\hat{q}_0} z)  \right )}
{\int {\rm D}t e^{-p {\cal Y}_0}} 
\right )^2 \right ]_{x^0,z} \cr
&&\phantom{q_1-q_0} \simeq \left [ \left (\frac{\partial^2 {\cal Y}_0^{\rm RS}}{\partial (\sqrt{\hat{q}_0}z )^2} \right )^2 
\left (\frac{\int {\rm D}t e^{-p {\cal Y}_0} t^2}{\int {\rm D}t e^{-p {\cal Y}_0}}
-\left (\frac{\int {\rm D}t e^{-p {\cal Y}_0} t}{\int {\rm D}t e^{-p {\cal Y}_0}} \right )^2  \right )\right ]_{x^0,z} (\hat{q}_1-\hat{q}_0) \cr
&&\phantom{q_1-q_0} \simeq \left [ \left (\frac{\partial^2 {\cal Y}_0^{\rm RS}}{\partial (\sqrt{\hat{q}_0}z )^2} \right )^2 
\right ]_{x^0,z} (\hat{q}_1-\hat{q}_0) 
\label{AT1}
\end{eqnarray}
and 
\begin{eqnarray}
&&\hat{q}_1-\hat{q}_0=2 \alpha \int {\rm D}z H\left (\frac{m}{\sqrt{\rho q-m^2}}z \right )
\left ( \frac{\int {\rm D}t e^{-p{\cal Y}_1} \left (
\partial{\cal Y}_1/\partial (\sqrt{{q}_0} z) \right )^2}
{\int {\rm D}t e^{-p {\cal Y}_1}} \right . \cr
&&\hspace*{4cm} \left . -
\left (\frac{\int {\rm D}t e^{-p {\cal Y}_1}
\left (\partial{\cal Y}_1/\partial (\sqrt{{q}_0} z)  \right )}
{\int {\rm D}t e^{-p {\cal Y}_1}} 
\right )^2 \right ) \cr
&&\phantom{\hat{q}_1-\hat{q}_0} \simeq 
2 \alpha \int {\rm D}z H\left (\frac{m}{\sqrt{\rho q-m^2}}z \right )
\left (\frac{\partial^2 {\cal Y}_1^{\rm RS}}{\partial (\sqrt{{q}_0}z )^2} \right )^2\cr
&& \hspace*{4cm} 
\times \left (\frac{\int {\rm D}t e^{-p {\cal Y}_1} t^2}{\int {\rm D}t e^{-p {\cal Y}_1}}
-\left (\frac{\int {\rm D}t e^{-p {\cal Y}_1} t}{\int {\rm D}t e^{-p {\cal Y}_1}} \right )^2  \right )
(q_1-q_0)\cr
&&\phantom{\hat{q}_1-\hat{q}_0} \simeq 
2 \alpha \int {\rm D}z H\left (\frac{m}{\sqrt{\rho q-m^2}}z \right )
\left (\frac{\partial^2 {\cal Y}_1^{\rm RS}}{\partial (\sqrt{{q}_0}z )^2} \right )^2 (q_1-q_0)
\label{AT2}
\end{eqnarray}
hold for $|q_1-q_0| \ll 1$ and $|\hat{q}_1-\hat{q}_0| \ll 1$ irrespectively of the value of $p$. Here ${\cal Y}_0^{\rm RS}$ and ${\cal Y}_1^{\rm RS}$ represent assessments of ${\cal Y}_0$ and ${\cal Y}_1$ under the assumptions of $\hat{q}_1=\hat{q}_0$ and $q_1=q_0$, respectively. In (\ref{AT1}) and (\ref{AT2}) we used the Taylor expansion expressions $\partial{\cal Y}_0/\partial (\sqrt{\hat{q}_0} z) \sim \partial{\cal Y}_0^{\rm RS} /\partial (\sqrt{\hat{q}_0} z) + \partial^2{\cal Y}_0^{\rm RS} /\partial (\sqrt{\hat{q}_0} z)^2 \sqrt{\hat{q}_1-\hat{q}_0} t$ and $\partial{\cal Y}_1/\partial (\sqrt{{q}_0} z) \sim \partial{\cal Y}_1^{\rm RS} /\partial (\sqrt{{q}_0} z) + \partial^2{\cal Y}_1^{\rm RS} /\partial (\sqrt{{q}_0} z)^2 \sqrt{q_1-q_0} t$, and the fact that the variances of $t$ for the measures ${\rm D}te^{-p {\cal Y}_0}/\int {\rm D}te^{-p {\cal Y}_0}$ and ${\rm D}te^{-p {\cal Y}_1}/\int {\rm D}te^{-p {\cal Y}_1}$ become unity as $\hat{q}_1 - \hat{q}_0$ and $q_1 -q_0$ vanish, irrespectively of the value of $p$.

To examine the stability of the RS solution in the limit of $\beta \to \infty$, let us change the variable notations as $\chi=\beta(1-q)$, $\hat{Q}+\hat{q}_1 \to \beta \hat{Q}$, $\hat{q}_1\to \beta^2 \hat{q}_1$, $\hat{q}_0 \to \beta^2 \hat{q}_0$, and $\hat{m}\to \beta \hat{m}$ and set $q_0=q$ and $\hat{q}_0=\hat{q}$. This yields expressions of ${\cal Y}^{\rm RS}_0 \simeq \beta \phi(\sqrt{\hat{q}}z+\hat{m}x^0; \hat{Q}) = -\beta g(\sqrt{\hat{q}}z+\hat{m}x^0)/\hat{Q}$ and ${\cal Y}_1^{\rm RS} \simeq (\beta/\chi) f(-\sqrt{q}z )$ for $\beta \gg 1$. Substituting these into (\ref{AT1}) and (\ref{AT2}) leads to 
\begin{eqnarray}
\Delta \simeq  \frac{1}{\hat{Q}^2} 
\left [ \left ( g^{\rm \prime \prime}(\sqrt{\hat{q}}z+\hat{m}x^0) \right )^2  \right ]_{x^0,z}
\hat{\Delta} 
\label{AT1new}
\end{eqnarray}
and
\begin{eqnarray}
\hat{\Delta} \simeq  \frac{2\alpha}{\chi^2}
\int {\rm D}z H \left (\frac{m}{\sqrt{\rho-m^2}} z\right )
\left ( f^{\prime \prime} (-z) \right )^2 \Delta,  
\label{AT2new}
\end{eqnarray}
where we set $\Delta=q_1-q$ and $\hat{\Delta}=\hat{q}_1-\hat{q}$, and used $q\to 1$. The condition that (\ref{AT1new}) and (\ref{AT2new}) allow a solution of $(\Delta,\hat{\Delta}) \ne (0,0)$ offers (\ref{AT}). 

\section{Derivation of the cavity equations}
\label{cavity_derivation}
We refer to the system in which $x_i$ and $(a_\mu, z_\mu)$ are kept out as the $i$-cavity and $\mu$-cavity systems, respectively. In addition, we denote ${\cal L}_{i \to \mu}(x_i)$, $A_{i \to \mu}$ and 
%% kaba 12/12/12
$H_{i \to \mu}$ 
as the single-body 
cost function for the $\mu$-cavity system and its parameters, respectively, 
and similarly for ${\cal L}_{\mu \to i}(a_\mu,z_\mu)$, 
%% kaba 12/12/12
$B_{\mu \to i}$ 
and $K_{\mu \to i}$. Self-consistent equations are derived from the following arguments. \\

\noindent{\bf Vertical step:} \\
Let us suppose that $x_i$ is put into the $i$-cavity system, which yields an approximation of the cost function of (\ref{coupled_cost}) as $(\Lambda/2)x_i^2 + |x_i| + \sum_{\nu =1}^M \left ({\cal L}_{\nu \to i}(a_\nu, z_\nu)+\Phi_{\nu i} a_\nu x_i \right )$. From this function we remove all terms that are related to $(a_\mu,z_\mu)$ of a certain index $\mu \in \{1,2,\ldots,M\}$, which leads to an approximate cost function of the $\mu$-cavity system. ${\cal L}_{i \to \mu}(x_i)$ must be obtained by partially optimizing the resulting $\mu$-cavity cost function with respect to 

$(a_\nu, z_\nu)$ 

of the remaining indices $\forall{\nu} \in \{1,2,\ldots,M\} \backslash \mu$, where $S \backslash a$ generally denotes the set provided by removing an element $a$ from a set $S$. This offers the relation
\begin{eqnarray}
{\cal L}_{i\to \mu}(x_i) = \frac{\Lambda}{2} x_i^2 + |x_i|+
\sum_{\nu \ne \mu} \left \{ \mathop{\rm min}_{z_\nu >0} \mathop{\rm max}_{a_\nu} 
\left \{{\cal L}_{\nu \to i}(a_\nu, z_\nu)+\Phi_{\nu i} a_\nu x_i \right \} \right \}.  
\label{Vstep}
\end{eqnarray}
This relation and the fact that $\Phi_{\mu i}$ is a negligibly small independent sample from an identical Gaussian distribution with zero mean and variance $N^{-1}$ yield the following equations evaluating $A_{i \to \mu}$ and $H_{i \to \mu}$ from a set of $\{B_{\nu \to i}\}$ and $\{K_{\nu \to i}\}$:
\begin{eqnarray}
A_{i \to \mu}&=&\Lambda+ \sum_{\nu \ne \mu} \frac{\Phi_{\nu i}^2}{B_{\nu \to i}} 
f^{\prime \prime} (K_{\nu \to i}), \label{VstepA} \\
H_{i \to \mu}&=& - \sum_{\nu \ne \mu} \frac{\Phi_{\nu i}}{B_{\nu \to i} }
f^{\prime }(K_{\nu \to i}). \label{VstepH}
\end{eqnarray}

\noindent{\bf Horizontal step:}\\
Similarly, putting $(a_\mu, z_\mu)$ into the $\mu$-cavity system and removing $x_i$ yields another relation,
\begin{eqnarray}
{\cal L}_{\mu \to i}(a_\mu, z_\mu)=- z_\mu a_\mu+\sum_{j \ne i} 
\left \{
\mathop{\rm min}_{x_j}\left \{
{\cal L}_{j \to \mu} (x_j)+\Phi_{\mu j} a_\mu x_j \right \} \right \}, 
\label{Hstep}
\end{eqnarray}
which offers 
\begin{eqnarray}
B_{\mu \to i}&=&\sum_{j \ne i} \frac{\Phi_{\mu j}^2}{A_{j \to \mu}} g^{\prime \prime}(H_{j \to \mu}), \label{HstepB} \\
K_{\mu \to i}&=&\sum_{j \ne i} \frac{\Phi_{\mu j}}{A_{j \to \mu}}g^{\prime}(H_{j \to \mu}).  \label{HstepK}
\end{eqnarray}
\\

\noindent{\bf Recovery step:}\\
$A_i$ and $H_i$ are evaluated from (\ref{HstepB}) and (\ref{HstepK}) as
\begin{eqnarray}
A_i&=&\Lambda+\sum_{\mu=1}^M \frac{\Phi_{\mu i}^2}{B_{\mu \to i}} 
f^{\prime \prime} (K_{\mu \to i}), \label{RstepA} \\
H_i&=& - \sum_{\mu=1}^M \frac{\Phi_{\mu i}}{B_{\mu \to i} }
f^{\prime }(K_{\mu \to i}).  \label{RstepH}
\end{eqnarray}
This means that the recovered signal is provided as
\begin{eqnarray}
\hat{x}_i =\frac{1}{A_i} g^\prime(H_i),  
\label{recoveredX}
\end{eqnarray}
where $\Lambda$ is determined in such a way that $\sum_{i=1}^N \hat{x}_i^2 =N$ holds. Similarly, 
\begin{eqnarray}
B_\mu&=& \sum_{i=1}^N \frac{\Phi_{\mu i}^2}{A_{i \to \mu}} g^{\prime \prime}(H_{i \to \mu}), \label{RstepB} \\
K_\mu&=& \sum_{i=1}^N \frac{\Phi_{\mu i}}{A_{i \to \mu}}g^{\prime}(H_{i \to \mu}), \label{RstepK} 
\end{eqnarray}
are obtained from (\ref{VstepA}) and (\ref{VstepH}). These offer the (approximate) optimal value of the Lagrange multiplier $a_\mu$ as
\begin{eqnarray}
\hat{a}_\mu=-\frac{1}{B_\mu} f^\prime(K_\mu). \label{recovereda}
\end{eqnarray}

Equations (\ref{VstepH}) and (\ref{RstepH}) indicate the difference between 

$H_{i \to \mu }$ and $H_{i}$ 

is vanishingly small for $N \to \infty$ as $\Phi_{\mu i}$ scales as $O\left (N^{-1/2} \right )$, and similarly for 

$K_{\mu \to i}$ and $K_\mu$. 

This also allows us to handle $A_i$ and $A_{i \to \mu}$ as a single site-independent parameter $A$, and we similarly deal with $B_\mu $ and $B_{\mu \to i}$ as $B$. These considerations, in conjunction with (\ref{RstepA}) and (\ref{RstepB}), offer 
\begin{eqnarray}
A&=&\Lambda+\frac{1}{NB} \sum_{\mu=1}^M f^{\prime \prime}(K_\mu ), 
\label{singleA}\\
B&=& \frac{1}{NA}\sum_{i=1}^N g^{\prime \prime}(H_i), 
\label{singleB}
\end{eqnarray}
where we replaced $\Phi_{\mu i}^2$ in (\ref{RstepA}) and (\ref{RstepB}) with its expectation $N^{-1}$ by utilizing the law of large numbers. Furthermore, inserting $f^{\prime}(K_{\mu \to i}) \simeq f^{\prime}(K_\mu-\Phi_{\mu i}\hat{x}_i) \simeq f^{\prime}(K_\mu)-\Phi_{\mu i} f^{\prime \prime}(K_\mu) \hat{x}_i$ and $g^{\prime}(H_{i \to \mu}) \simeq g^{\prime}(H_i+ \Phi_{\mu i} \hat{a}_\mu ) \simeq g^{\prime}(H_i)+\Phi_{\mu i} g^{\prime \prime}(H_i) \hat{a}_\mu$ into (\ref{RstepH}) and (\ref{RstepK}), respectively, yields
\begin{eqnarray}
H_i &\simeq&\sum_{\mu=1 }^M \Phi_{\mu i} \hat{a}_\mu + \left (\frac{1}{B}\sum_{\mu=1}^M 
\Phi_{\mu i}^2 f^{\prime \prime}(K_\mu) \right ) \hat{x}_i \cr
&\simeq &\sum_{\mu=1 }^M \Phi_{\mu i} \hat{a}_\mu + \left (\frac{1}{NB}\sum_{\mu=1}^M 
 f^{\prime \prime}(K_\mu) \right ) \hat{x}_i 
\cr
&=& \sum_{\mu=1 }^M \Phi_{\mu i} \hat{a}_\mu + \Gamma \hat{x}_i 
\label{singleH}
\end{eqnarray}
and
\begin{eqnarray}
K_\mu &\simeq& \sum_{i=1}^N \Phi_{\mu i} \hat{x}_i -\left (\frac{1}{A}\sum_{i=1}^N \Phi_{\mu i}^2 
g^{\prime \prime} (H_i) \right )\hat{a}_\mu \cr
&\simeq & \sum_{i=1}^N \Phi_{\mu i} \hat{x}_i -\left (\frac{1}{N A}\sum_{i=1}^N  
g^{\prime \prime} (H_i) \right )\hat{a}_\mu \cr
&=& \sum_{i=1}^N \Phi_{\mu i} \hat{x}_i -B \hat{a}_\mu , 
\label{singleK}
\end{eqnarray}
where we set $\Gamma=(NB)^{-1}\sum_{\mu=1}^M f^{\prime \prime}(K_\mu)$. Equations (\ref{recoveredX}), (\ref{recovereda}), and (\ref{singleA})--(\ref{singleK}) lead to (\ref{BPnaive1})--(\ref{BPnaive4}).

\end{document}